# Quantum Computing

Linear optics implementations

*Pål Sundsøy and Egil Fjeldberg*

"The nineteenth century was known as the machine age, the twentieth century will go down in history as the information age. I believe the twenty-first century will be the quantum age."

— *Paul Davies (1996)*

"..it seems that the laws of physics present no barrier to reducing the size of computers until bits are the size of atoms, and quantum behavior holds sway"

— *Nobel Prize winner physicist Richard Feynman (1985)*

## Preface

This work was carried out as a directed study at the department of Physics at the Norwegian University of Science and Technology under the guidance of Prof. Johannes Skaar from the Department of Physical Electronics.
We would especially like to thanks Prof. Skaar for his assistance and source of knowledge on the subject.
Also, we would like to thank Prof. Bo-Sture Skagerstam at the Department of Physics who has been a good source of inspiration within the field of Quantum Optics.

Pål Roe Sundsøy and Egil Fjeldberg
*NTNU, Trondheim*
*December 2003*

# Contents

# List of Figures

# List of Tables

## Abstract

One of the main problems that optical quantum computing has to overcome is the efficient construction of two-photon gates. Theoretically these gates can be realized using Kerr-nonlinearities, but the techniques involved are experimentally very difficult. We therefore employ linear optics with projective measurements to generate these non-linearities. The downside is that the measurement-induced nonlinearities achieved with linear optics are less versatile and the success rate can be quite low.
This project is mainly the result of a literature study but also a theoretical work on the physics behind quantum optical multiports which is essential for realizing two-photon gates. Applying different postcorrection techniques we try to increase the probability of success in a modified non-linear sign shift gate which might be foundational for the two photon controlled-NOT gate. We have proven that it's not possible to correct the states by only using a single beam splitter.However, it might be possible to increase the probability of success using a more complex setup with at least two error-correcting beam splitters.

# Chapter 1

# Introduction

## 1.1 Background

Civilizations has advanced as people discovered new ways of exploiting various physical resources such as materials, forces and energies. In the twentieth century information was added to the list when the invention of computers allowed complex information processing to be performed outside human brains.

Computer users have been used to an exponential increase in computing speed and capacity over the past few decades. In 1965 Gordon Moore observed that the number of transistors per square inch on integrated circuits had doubled every year since the integrated circuit was invented. In subsequent years, the pace slowed down a bit, but data density has doubled approximately every 18 months, and this is the current definition of Moore's law. But even Moore's law has limits.

For each new chip generation, the doubling of capacity means that about half as many atoms are being used per bit of information. When we project this into the future, this trend reaches a limit of one atom per bit of information sometime between 2010 and 2020. This will not necessarily slow down the improvements of computing. One new technology, quantum computing, has the potential to dramatically to increase the rate of advances in computational power. A key future of a quantum computer is that it can go beyond a classical Turing machine. That is, there are functions that can be computed on a quantum computer that cannot be effectively computed with a conventional computer.

In 1982 the Nobel Prize winner physicist Richard Feynman was trying to simulate the interaction of a quantum mechanical system with $N$ particles using a classical turing machine [Fey85]. Try as he might, he was unable to find a general solution without using exponential resources.
At the same time, David Deutsch tried to create the most powerful model of computation using the laws of physics. In 1985 he made the notion of a Universal Quantum Computer based on the laws of quantum mechanics [Deu85]. He also suggested that quantum computers may be more powerful than classical computers.
The next breaktrough came in 1994 when Peter Shor demonstrated that quantum computers could efficiently factor large numbers [Sho85]. This was of great interest since it is widely believed that no efficient factoring algorithm is possible for classical computers.
However, there are no roses without thorns. It was initially unclear whether quantum computation (QC) was a physically realizable model. The fact that in nature, quantum effects are rarely observable, and in fact physical noise processes rapidly remove the necessary phase relationships. To solve the problem of quantum noise, P.Shor and A.Steane introduced quantum error-correcting codes [Ste96b]. This idea was that under physically reasonable assumptions, fault tolerant quantum computation is possible.
As a result there are now many intense experimental efforts devoted toward realizing quantum computation, in a wide and increasing variety of physical systems.

## 1.2 Applications for Quantum Computers

Quantum computers can do everything a classical computer can do - and more. But to date there have been only two major algorithms for quantum computers which are significantly better than for classical computers. Shor's factorization and Grover's datasearch, but it's believed that there are a number of killer applications which will enjoy exponential speed-ups over classical computation while requiring only a small or modest number of qubits[1].

### 1.2.1 Shor's algorithm

Peter Shor's 1994 discovery of an efficient quantum algorithm for finding the prime factors of large integers is the biggest success so far. Shor's algorithm takes advantage of quantum parallelism and finds a factor of a number $N$ in time roughly the square of the length of the input (which is $\log N$ bits). In contrast, every known classical algorithm requires exponential time to factor. Since public-key cryptography, notably the RSA-system rely on factorization of large integers, the current cryptography becomes totally insecure if quantum computers come to life.

### 1.2.2 Quantum Cryptography

In 1984, Bennet and Brassard found a scheme which allowed two distant parties to obtain a shared secret key via quantum mechanical communication [BBE92]. Their scheme was always believed to be fully secure against any type of eavesdropper, and recently this has been proven. On the other hand some other parts of electronic transactions, like unforgeable signatures, appear to be beyond the power of quantum methods.

### 1.2.3 Grover's algorithm

Grover's 1996 algorithm for searching databases is a third application [Gro96]. Consider finding some specific cd in a large unordered database of $N$ items. Classically, there is no better method than just go through all records sequentially, which requires expected $\frac{N}{2}$ time steps for cd in general position. However Grover's algorithm uses quantum superpositions to examine all cd's "at the same time", and finds the desired cd in roughly $\sqrt{N}$ steps.

[1] A qubit is the fundamental unit in a quantum computer. Unlike classical bits qubits can be in a linear superposition of states $|0\rangle$ or $|1\rangle$, until measurement.The general expression for a qubit is $|\psi\rangle = \alpha|0\rangle + \beta|1\rangle$, with $|\alpha^2| + |\beta|^2 = 1$

## 1.3 Physical requirements for QC

How to create a physical device that operate in the unexplored quantum mechanical regime is still a challenge. There are for the time being done many efforts in creating models, and there already exists some primitive working models.

In 2000 DiVencenzo proposed five general implementation-requirements, plus two relating to the communication of quantum information [DiV00]. We allow us to reproduce the requirements.

**A scalable physical system with well characterized qubits.** Its physical parameteres should be accurately known, including the internal Hamiltonian of the qubit, the presence of the couplings to other states of the qubit, interaction with other qubits and couplings to external fields that might be used to manipulate the state of the qubit.

**The ability to initialize the state of the qubits to a simple fiducial state.** The register should be initialized to a known value before the start of computation.

**Long relevant decoherence[2] times, much longer than the gate operation time.** Decoherence is very important for the fundamental of quantum physics, as it is identified as the principal mechanism for emergence of classical behavior.

**A universal set of quantum gates.** To allow any kind of information processing preferably with a small number of resources

**A qubit-specific measurement capability,** to be able to read out the result of computation.

**The ability to interconvert stationary flying qubits.** Qubits can be taken into a form easily transmitted (e.g. photons in optical fibers)

**The ability faithfully to transmit flying qubits between specified locations.** The state must be preserved during the transmission.

[2]Decoherence times characterize the dynamics of a qubit in contact with its environment. The simplified definition of this time is that it is the characteristic time for a generic qubit state $|\psi\rangle = \alpha|0\rangle + \beta|1\rangle$ to be transformed into the mixture $\rho = |\alpha|^2|0\rangle\langle 0| + |\beta|^2|1\rangle\langle 1|$

### 1.3.1 Constructing qubits in physical systems

Building a quantum computer, with more than a trivial number of qubits, is a problem plaguing the researchers today. Realizing a quantum computer will require the ability to manipulate significantly more qubits than we currently have the technology to control today.
A wide variety of possible implementations have been investigated. We will beneath list some of the proposals:

**Cavity QED** where the qubits are usually stored in Zeeman ground states of trapped atoms, fixed at regular intervals in an optical cavity [PGCZ00]. Each atom maybe individually 'addressed' with it's own laser. The states of single atoms are influenced using laser pulses and the states of atoms can be entangled by the exchange of a cavity photon. The main sources of decoherence for this type of set-up are those caused by spontaneous emission and those caused by cavity decay during gate operation.

**Trapped Ions** where the qubit are stored in ions, the two states correspond to two of the internal states of the trapped ion which are confined in a linear trap [CZ95, MMKI95]. Laser beams also play a role in performing operations in this system and can be used to independently manipulate single qubits. Unitary transformations on $n$ qubits can be accomplished by exciting the collective quantized motion of the ions with lasers. Sources of decoherence in this type of method are spontaneous decay in the internal atomic states of the ions and also damping in the motion of the ions.

**Bulk Spin Resonance** is completely different from the above methods, using common liquids and radio frequency pulses from Nuclear Magnetic Resonance spectroscopy methods. This method represents a more 'hands on' method of testing quantum algorithms (in some limited fashion) [VSBC00].

Macroscopic ensembles are used to store qubit states, which are uniquely protected by 'ordering' the majority of the molecules in the sample thus providing a protective coat for a small number of molecules that can 'hold' a pure quantum state.

Using traditional pulsed NMR methods and some novel excitation processes unitary transformations can be accomplished.

However using this technique requires some modification to existing algorithms, but using sufficiently complex molecules suspended in liquid possibly 20-30 qubit computations can be done.

**Linear Optics** implementations are a progressing field, where qubits are represented by location of single photon modes or polarization. The problem so far have been the need for non-linear couplings among photons, which is a necessary requirement for optical quantum computation in general [LB99]. However, by use of feedback from the result of photodetectors, which we will study in later chapters, it seems that we can replace a nonlinear component with it.

## 1.4 Outline

This report will be divided in 9 Chapters and an appendix containing our Maple calculations.

We start by summarizing the most important happenings within quantum computing, before we continue in Chapter 3 with the notation used in rest of the report together with the theory behind optical components used in linear optics quantum computing (LOQC).
In Chapter 4 we talk about the various ways of representing a bosonic qubit, followed by Chapter 5 which attach importance to the accomplishments in LOQC. Chapter 6 is mainly a brief introduction to the optical methods which we in some sense will take advantage of in Chapter 7 and 8.
Chapter 7 explains the use of postselection, while we in Chapter 8 try to find out how the use of postcorrection can increase the probability of a successful result in a quantum gates. At the end in this chapter we also take a look at what's been accomplished within the very important controlled NOT-gate (CNOT).
In Chapter 9 we give a short discussion and conclusion.

# Chapter 2

# Timeline of quantum computing

## 2.1 A brief history of quantum computing

In spite of that quantum computing is a young and developing field, both theoretically and experimentally, it is important to know the story and see the development from an aerial point of view.
Beneath we have listed some important happenings, and emphasized the photonic breakthroughs.

**1981** Richard Feynman gave the first proposal for using quantum phenomena to perform computations. The speech was entitled "Simulating Physics With Computers". It was in a talk he gave at the First Conference on the Physics of Computation at MIT.

**1985** David Deutsch, at the University of Oxford, described the first universal quantum computer. Just as a universal Turing machine can simulate any other Turing machine efficiently, so the universal quantum computer is able to simulate any other quantum computer with at most a polynomial slowdown. This raised the hope that a simple device might be able to perform many different quantum algorithms.

**1988** *Milburn propose a simple optical model to realize a reversible Fredkin gate. The device makes use of the Kerr nonlinearity to produce intensity dependent phase shifts. Unfortunately this gate need non-linear couplings that is too large for today's technology* [Mil88].

**1994** Peter Shor, at ATT's Bell Labs in New Jersey, discovered a remarkable factoring algorithm which could theoretically break many of the cryptosystems in use today. Its invention sparked a tremendous interest in quantum computers, even outside the physics community.

**1995** Shor proposed the first scheme for quantum error correction which could be a key technology for building large-scale quantum computers that work. Active research on this continues.

**1995** *A. Turchette al. exploits the optical nonlinearities realizable in cavity electrodynamics. Spontaneous emission and coupling of the photon in/out of the cavity is a problem* [THL95].

**1996** Lov Grover, at Bell Labs, invented the quantum database search algorithm.

**1997** *Cerf,Adami and Kwait present a systemathic method for simulating small-scale quantum circuits by use of linear optical devices* [CAK98].

**1997** David Cory, A.F. Fahmy and Timothy Havel, and at the same time Neil Gershenfeld and Isaac Chuang at MIT published the first papers on quantum computers based on bulk spin resonance, or thermal ensembles.

**1998** First working 2-qubit NMR computer demonstrated at University of California Berkeley.

**1999** First working 3-qubit NMR computer demonstrated at IBM's Almaden Research Center. First execution of Grover's algorithm.

**2000** First working 5-qubit NMR computer demonstrated at IBM's Almaden Research Center. First execution of order finding (part of Shor's algorithm).

**2001** First working 7-qubit NMR computer demonstrated at IBM's Almaden Research Center. First execution of Shor's algorithm. The number 15 was factored using 1018 identical molecules, each containing 7 atoms.

**2001** *Knill, Laflamme and Milburn shows that efficient quantum computation is possible using only linear optical elements, single photon sources and photodetectors* [KLM01].

# Chapter 3

# General formalism

*- The principles of quantum computing are posed using the most fundamental ideas of quantum mechanics, ones whose embodiment can be contemplated in virtually every branch of quantum physics.*

## 3.1 Notation

In this report we will make use of the standard Dirac notation [Dir47] for the states and operators. The states are represented by "ket" vectors $|\psi\rangle$, or alternatively by a "bra"-vector $\langle\psi|$. The vector $|\psi\rangle$ is said to describe the state of the system. According to the superposition principle in quantum physics, the set of states form a linear vector space over the complex numbers and is denoted by $\mathcal{H}$. By definition

$$|\psi\rangle^{\dagger} = \langle\psi|, \tag{3.1}$$

where $|\psi\rangle^{\dagger}$ denotes the hermitian conjugate of $\langle\psi|$. The innerproduct for e.g. $|\phi\rangle$ and $|\psi\rangle$ will be denoted by $\langle\phi|\psi\rangle$

We will work with $n$ qubits, the state of which is a unit vector in the complex Hilbert space $\mathcal{C}^2 \otimes \mathcal{C}^2 \otimes ... \otimes \mathcal{C}^2$.As the natural basis for this space, we take the basis consisting of $2^n$ vectors:

$$\begin{array}{c} |0\rangle \otimes |0\rangle \otimes .... \otimes |0\rangle \\ |0\rangle \otimes |0\rangle \otimes .... \otimes |1\rangle \\ \vdots \\ |1\rangle \otimes |1\rangle \otimes .... \otimes |1\rangle \end{array} \tag{3.2}$$

For brevity, we will omit the tensor product, and denote

$$|i_1\rangle \otimes |i_2\rangle \otimes .... \otimes |i_n\rangle = |i_1, i_2, ..., i_n\rangle \equiv |i\rangle \tag{3.3}$$

where $i_1, i_2, ...i_n$ is the binary representation of the integer i, a number between 0 and $2^n - 1$.
The matrices of the operators will be expressed with capital letters.

## 3.2 Quantum Computation

According to quantum mechanics, the time development of any quantum mechanical system is governed by the Schrödinger equation

$$i\hbar \frac{d}{dt}|\psi\rangle = H|\psi\rangle, \tag{3.4}$$

In fact we know that this is only a suitable approximation for every day speeds and energies. At high speeds and energies we must use the Dirac equation and Einstein relativity, with its predictions of relativistic mass increase and particle-antiparticle creation, must be taken into account. However the standard non-relativistic quantum mechanics will suffice, and in accordance to the Schrödinger equation any kind of operation on a qubit system is given in the form of unitary operators

$$U = e^{-\frac{i}{\hbar}\int H dt}, \tag{3.5}$$

where $H$ is the regular Hamilton operator as defined in quantum mechanics. From the constraint of unitarity it immediately follows that all operations performed on a qubit system must be reversible. That is, the reverse time evolution specified by the unitary operator $U^{-1} = U^{\dagger}$ always exists.

Quantum computation is usually described by unitary operations because time evolution of a closed system is described by unitary transformations. The basic idea that a useful computer might be constructed which operates according to the principle of unitary evolution was first put forward by Benioff [Ben82].

### 3.2.1 Quantum Gates

Quantum Gates are the building blocks in quantum computation analogous to digital gates in electronic digital computers. Simple unitary operations on qubits are called quantum logic gates [Deu85, Deu89].

These gates alter the states and thereby manipulate the quantum information. A set of quantum gates and wires makes a quantum circuit.
In general an arbitrary quantum computation on any number of qubits can be generated by a finite set of gates that is said to be *universal* for quantum computation. Or explained in a different way - a set of gates is said to be universal for quantum computation if any unitary operation may be approximated to arbitrary accuracy by a quantum circuit involving only those gates. A quantum gate must also fullfill the reversibility criteria.

### Single Qubit Gates

Quantum gates on a single qubit can be described by unitary $2 \times 2$ matrices. Some of the most important one-qubit gates are the pauli matrices:

$$X \equiv \begin{pmatrix} 0 & 1 \\ 1 & 0 \end{pmatrix} \quad Y \equiv \begin{pmatrix} 0 & -i \\ i & 0 \end{pmatrix} \quad Z \equiv \begin{pmatrix} 1 & 0 \\ 0 & -1 \end{pmatrix} \tag{3.6}$$

As an example, we operate with the Pauli $X$-matrix on the single qubit state[1]
$|\psi\rangle = \alpha|0\rangle + \beta|1\rangle = \alpha \begin{pmatrix} 1 \\ 0 \end{pmatrix} + \beta \begin{pmatrix} 0 \\ 1 \end{pmatrix} = \begin{pmatrix} \alpha \\ \beta \end{pmatrix}$, then

$$X|\psi\rangle = \begin{pmatrix} 0 & 1 \\ 1 & 0 \end{pmatrix} \begin{pmatrix} \alpha \\ \beta \end{pmatrix} = \begin{pmatrix} \beta \\ \alpha \end{pmatrix} = \alpha|1\rangle + \beta|0\rangle \tag{3.7}$$

We see that the action of the gate is to switch the probability amplitudes of the states.

In general, an arbitrary single qubit gate can be decomposed as a product of rotations [NC00]

$$U = e^{i\alpha} R_x(\beta) R_y(\gamma) R_z(\delta) \tag{3.8}$$

where $R_x(\beta)$,$R_y(\gamma)$ and $R_z(\delta)$ are rotation operators around $\hat{x}, \hat{y}$ and $\hat{z}$ axes, $\beta, \gamma$ and $\delta$ real numbers and $e^{i\alpha}$ represent a global phase shift. This decomposition can be used to give an exact prescription for performing an arbitrary single qubit quantum logic gate.

---

[1] Here we have chosen the computational basis for the qubit $\Rightarrow$ $|0\rangle \equiv \begin{pmatrix} 1 \\ 0 \end{pmatrix}$ and $|1\rangle \equiv \begin{pmatrix} 0 \\ 1 \end{pmatrix}$

The rotation operator around the $\hat{x}$ axis is defined to

$$\begin{aligned} R_x(\theta) \equiv e^{\frac{-i\theta X}{2}} &= \sum_{n=0}^{\infty} \frac{(\frac{-i\theta\hat{X}}{2})^n}{n!} \\ &= \mathbb{I} \sum_{\text{equal n}} \frac{(\frac{-i\theta}{2})^n}{n!} + X \sum_{\text{odd n}} \frac{(\frac{-i\theta}{2})^n}{n!} \\ &= \mathbb{I} \cos\frac{\theta}{2} - iX \sin\frac{\theta}{2} \\ &= \begin{pmatrix} \cos\frac{\theta}{2} & -i\sin\frac{\theta}{2} \\ -i\sin\frac{\theta}{2} & \cos\frac{\theta}{2} \end{pmatrix} \end{aligned} \tag{3.9}$$

where we have made use of $X^2 = \mathbb{I}$ Following the same procedure $R_y(\theta)$ and $R_z(\theta)$ can be found to be

$$R_y(\theta) = \begin{pmatrix} \cos\frac{\theta}{2} & -\sin\frac{\theta}{2} \\ \sin\frac{\theta}{2} & \cos\frac{\theta}{2} \end{pmatrix} \tag{3.10}$$

$$R_z(\theta) = \begin{pmatrix} e^{\frac{-i\theta}{2}} & 0 \\ 0 & e^{\frac{i\theta}{2}} \end{pmatrix} \tag{3.11}$$

It may be shown that all one-qubit rotations can be implemented with linear optics [KLM01] which will be understood in Chapter 3.3.5.

### Multiple Qubit Gates

The CNOT, also known as the XOR or measurement gate is a 2-qubit gate we will study in more detail.

This gate takes two qubits as input known as the control - and target qubit, $|x\rangle$ and $|y\rangle$ respectively. If the value of $|x\rangle$ is 0, $|y\rangle$ remains the same; otherwise the value of $|y\rangle$ is flipped to its opposite. Summarized the action of the gate can be described as $|x, y\rangle \rightarrow |x, x \oplus y\rangle$ where $\oplus$ is addition modulo two.

In the basis $\{|x\rangle|x\rangle, |x\rangle|y\rangle, |y\rangle|x\rangle, |y\rangle|y\rangle\}$ the CNOT gate, $U_{CNOT}$, acting on 2 qubits can be written as a simple matrix:

$$\mathrm{U_{CNOT}}|\psi\rangle = \overbrace{\begin{pmatrix} 1 & 0 & 0 & 0 \\ 0 & 1 & 0 & 0 \\ 0 & 0 & 0 & 1 \\ 0 & 0 & 1 & 0 \end{pmatrix}}^{U_{CNOT}} \begin{pmatrix} \alpha \\ \beta \\ \gamma \\ \delta \end{pmatrix} = \begin{pmatrix} \alpha \\ \beta \\ \delta \\ \gamma \end{pmatrix} \quad (3.12)$$

This gate is of special interest because in combination with single qubit gates it can constitute any two-qubit gates, which is universal. That is , any quantum logic gate may be built from CNOT and single qubits gates [BDEJ95].

## 3.3 Theory of linear multiports

Linear multiports are a generalization of some of the most important optical elements; beam splitters and phase shifters. A beam splitter can be viewed as a 4-port device, a black box with two inputs and two outputs. The operation of such a device can be formally described by a unitary transformation of the two input modes into the two output modes.

Before going on considering different linear optics implementations we need a mathematical toolkit from the field of quantum optics and we need to know how to apply this kit on linear multiports.

In general a quantum system with $N$ discrete states is described by a vector $|\psi_0\rangle$ in a N-dimensional Hilbert space. This vector can be transformed into another vector $|\psi_1\rangle$. The transformation can be described by a matrix $U$ and this transformation matrix must conserve probability for a lossless system:

$$||\psi_1\rangle|^2 = |U|\psi_0\rangle|^2. \quad (3.13)$$

This condition is fulfilled if the matrix is unitary, that is the inverse of the matrix is equal to its transposed complex conjugate:

$$U^{-1} = (U^*)^T \quad (3.14)$$

Thus unitary matrices describe the transformation between states in the discrete Hilbert space.

### 3.3.1 Fock states

The bosonic qubits we are about to use are defined by states of optical modes. The time evolution of a cavity mode of electromagnetic radiation is modelled quantum mechanically by a harmonic oscillator. By quantization of the electromagnetic field [Lou83] we obtain the well-known operator properties,

$$a|n\rangle = \sqrt{n}|n-1\rangle \tag{3.15}$$
$$a^\dagger|n\rangle = \sqrt{n+1}|n+1\rangle \tag{3.16}$$

where $a$ is the destruction operator and $a^\dagger$ is the creation operator, and $n = 0, 1, ...$ is the number of photons in the mode. We also note that $a|0\rangle \equiv 0$ since there is no eigenstate of lower energy than the groundstate, $n \geq 0$.

To generate any Fock state of the $k$th mode $|n_k\rangle$ from the vacuum , we apply (3.15) several times, getting

$$|n_k\rangle = \frac{(a_k^\dagger)^{n_k}}{\sqrt{n_k!}}|0\rangle \tag{3.17}$$

The eigenstates $|n\rangle$ are also often referred to as photon number states. They form a complete set and they are orthogonal, i.e.,

$$\sum_{n=0}^{\infty} |n_k\rangle\langle n_k| = 1$$
$$\langle n_k|m_k\rangle = \delta_{nm} \tag{3.18}$$

### 3.3.2 Linear optical multiports

Linear optics transformations that preserve the photon number are called passive linear.
A linear quantum gate with $N$ input ports [Ska00], visualized in Fig. 3.1, can be written in the classical sense as a relation between outgoing and incoming waves

$$b_i = \sum_{j=1}^{N} \Lambda_{ij} a_j \tag{3.19}$$

where $b_i$ $(a_i)$ denotes the complex amplitudes of the outgoing (incoming) waves, and $\Lambda_{ij}$ is a unitary transformation matrix from port $i$ to port $j$ which depends on the linear optics element we choose.

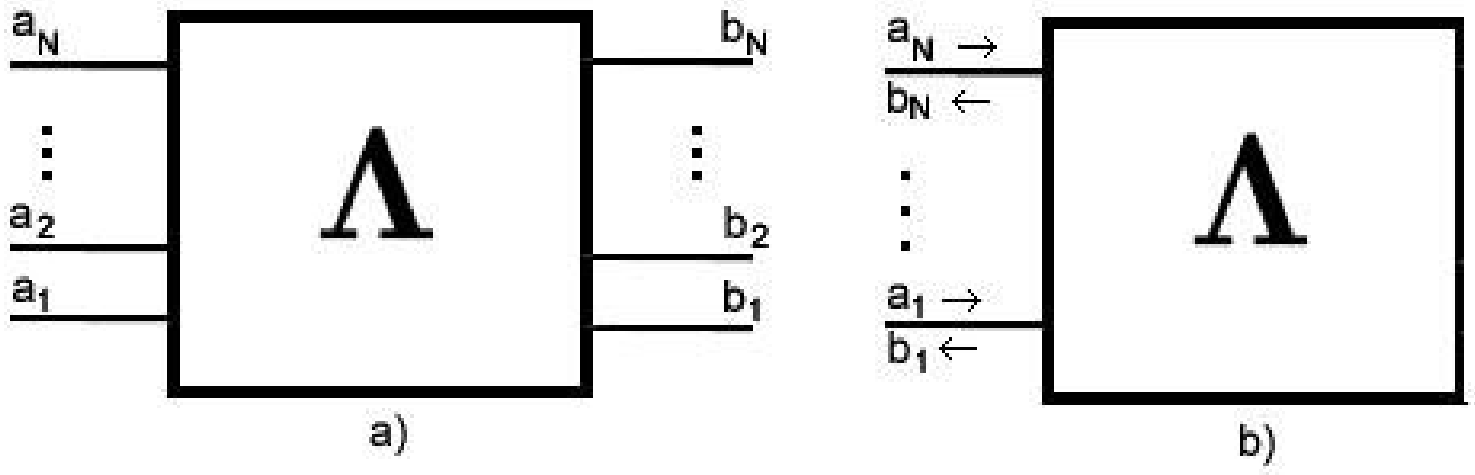

**Figure 3.1:** *Linear quantum gate consisting of N input ports and N output ports.(a) Different physical ports (b) Coincident ports.*

In the quantum mechanical case $a_i$ and $b_i$ become operators which satisfy the commutation relations

$$[a_i, a_j^\dagger] = \delta_{ij} \ , \quad [b_i, b_j^\dagger] = \delta_{ij} \ , \quad [a_i, a_j] = 0] \ , \quad [b_i, b_j] = 0 \tag{3.20}$$

If we now consider $N$ input ports we can define a basis which can be written as a product of number states for each mode

$$|n_N\rangle \otimes ... \otimes |n_2\rangle \otimes |n_1\rangle = |n_N...n_2n_1\rangle. \tag{3.21}$$

There is absolutely no way in which one can determine which path a photon has taken inside a multiport device, we must add up the phases accumulated over all paths inside the systems and then add the probability amplitudes of all the possibilities. This transforms to a superposition of terms on the form $|m_N...m_2m_1\rangle$ after the transformation matrix, $\Lambda$, has operated. If we consider a lossless system the photon number will be preserved under this transformation, $\sum_{k=1}^{N} n_k = \sum_{k=1}^{N} m_k$.

The operators, in accordance to the Heisenberg-picture[2] will evolve as

[2]The Heisenberg picture of quantum mechanics absorbs the time dependence into the operator, keeping the basis fixed. The Schrödinger picture, on the other hand, keeps the operator fixed in time but changes the basis. An arbitrary operator has to evolve as $A \rightarrow U^\dagger A U$ where $U$ is the unitary evolution operator that preserves the particle number.

$$a_i^\dagger \rightarrow b_i^\dagger = \sum_{k=1}^{N} \Lambda_{ik}^* a_k^\dagger \equiv U^\dagger a_i^\dagger U \tag{3.22}$$

$$a_i \rightarrow b_i = \sum_{k=1}^{N} \Lambda_{ik} a_k \equiv U^\dagger a_i U \tag{3.23}$$

$$\tag{3.24}$$

Since $\Lambda_{ik}$ is unitary we can rewrite (3.22) to

$$U a_i^\dagger U^\dagger = \sum_{k=1}^{N} \Lambda_{ki} a_k^\dagger \tag{3.25}$$

which will be useful in our further analysis.

Now, given an arbitrary input for the $N$th mode the number state can be written as

$$|n_N..n_2 n_1\rangle = \frac{a_N^{\dagger n_N} ... a_2^{\dagger n_2} a_1^{\dagger n_1}}{\sqrt{n_N! ... n_2! n_1!}} |0...0\rangle \tag{3.26}$$

in accordance to (3.18) and (3.21).
To find the exit-state we operate on this number state with a unitary operator $U$ and keep in mind that

$$U a_i^{\dagger 2} = U a_i^\dagger \mathbb{I} a_i^\dagger \mathbb{I} = U a_i^\dagger U^\dagger U a_i^\dagger U = (U a_i^\dagger U^\dagger)^2 U \tag{3.27}$$

which can be repeated as many times as necessary to obtain the operator that comes from the evolution under U of a generic $a_i^{\dagger n_i}$. This gives

$$\begin{aligned} U|n_N..n_2 n_1\rangle &= \frac{(U a_N^\dagger U^\dagger)^{n_N} ... (U a_2^\dagger U^\dagger)^{n_2} (U a_1^\dagger U^\dagger)^{n_1}}{\sqrt{n_N! ... n_2! n_1!}} U|0...0\rangle \\ &= \frac{\prod_{k=1}^{N} (\sum_{l=1}^{N} \Lambda_{lk} a_l^\dagger)^{n_k}}{\sqrt{n_N! ... n_2! n_1!}} |0...0\rangle \end{aligned} \tag{3.28}$$

where we have made use of (3.25) and taken into account that the vacuum state remains the same under unitary evolution, $U|0\rangle = |0\rangle$.

This equation becomes handy when we later will calculate probabilities for detecting photons in given output modes when the input modes and transformation matrix, $\Lambda$, is given.

It may be shown that for every unitary $\Lambda$, the corresponding $U$ is implementable by a network of beam splitters and phase shifters [ZRBB94].

### 3.3.3 General beam-splitter tranformation

A beam splitter has two input ports and two output ports, as shown in Fig. 3.2. In classical optics, input ports which experience no intensity may be neglected. In the quantum optical case we must consider both input ports at all times. The vacuum state ($|0\rangle$) is also a state of the electromagnetic field and may interfere with the other input state. Every lossless beamsplitter [YMK86, PSM87, RCT89] can be thought of as a unitary operator on the level of photonic creation and annihilation operators of the incoming and outgoing fields, i.e.

$$b = U^{\dagger} a U = \Lambda a \ , \ a = \begin{pmatrix} a_1 \\ a_2 \end{pmatrix} \ , \ b = \begin{pmatrix} b_1 \\ b_2 \end{pmatrix} \tag{3.29}$$

where $U$ is a unitary operator and $\Lambda$ the associated unitary matrix($\Lambda \in SU(2)^3$). This transformation matrix consist of the transmission and reflection coefficients $T$ and $R$ and can be given in the form

$$\Lambda = \begin{pmatrix} R & T \\ T^* & R^* \end{pmatrix} = \begin{pmatrix} \sqrt{\eta} & \sqrt{1-\eta} \\ \sqrt{1-\eta} & -\sqrt{\eta} \end{pmatrix} \tag{3.30}$$

where the transmittivity is given by $T = 1 - \eta$ and the reflectivity is given by $R = \eta$.

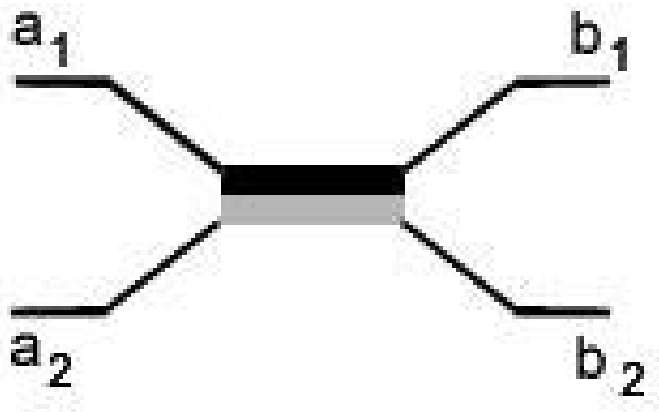

**Figure 3.2:** *Sketch of a asymmetrical beamsplitter. Gray indicates the surface from which a sign change occurs upon reflection.*

We require that $|T|^2 + |R|^2 = 1$ because of the unitarity of $\Lambda$. This leads to the definition of the beam splitter's 'angle' by writing $|T| = \cos\varphi$ and

[3] The group $SU(N)$,N=2,3.. consists of complex $N \times N$ matrices U obeying $U^{\dagger}U = 1$ , $UU^{\dagger} = 1$ , $\det U = 1$

$|R| = \sin\varphi$.
It can be easily seen from Fig. 3.2 that

$$\begin{aligned} b_1 &= \sqrt{\eta}a_1 + \sqrt{1-\eta}a_2 \\ b_2 &= \sqrt{1-\eta}a_1 - \sqrt{\eta}a_2 \end{aligned} \tag{3.31}$$

which is equivalent to (3.29).

### 3.3.4 General phase-shifter tranformation

The simplest optical element,the phase shifter $P$, acts like normal time evolution, but at a different rate, and only affects the modes going through it. If we act with a phase shifter on a vacuum state, $P|0\rangle = |0\rangle$, nothing happens. On the other hand, if we act on a single photon state one obtains $P|1\rangle = e^{i\Delta}|1\rangle$ where $\Delta \equiv \frac{(n-n_0)L}{c_0}$ is the difference between propagation times of light in vacuum and the medium. The transformation matrix for the system of Fig. 3.3 is

$$\Lambda = \begin{pmatrix} e^{i\triangle} & 0 \\ 0 & 1 \end{pmatrix} \tag{3.32}$$

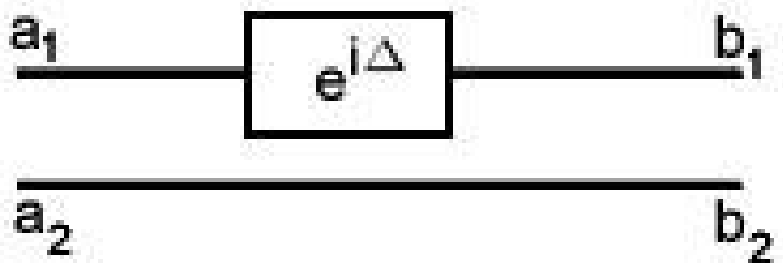

**Figure 3.3:** *Sketch of a phase shifter. A wave entering in $a_1$ will be delayed with respect to one entering in $a_2$.*

### 3.3.5 Z-Y decomposition on a single qubit using phase shifters and beam splitters

To show that one-qubit rotations can be implemented with linear optics, we will show that $R_y(\gamma)$ and $R_z(\beta)$ (from Eq.(3.10) and Eq.(3.11)) can be realized using beam splitters and phase shifters. These two qubit-rotations are sufficient for performing an unitary operation $U$ on a single qubit,

$$U = e^{i\alpha}R_z(\beta)R_y(\gamma)R_z(\delta) \tag{3.33}$$

Here we have replaced $R_x(\beta)$ with $R_z(\delta)$ from Eq.(3.8) which means that we rotate twice around the $\hat{z}$-axis to get the $U$ we want.

By introducting the parametrization $\eta = \cos^2 \frac{\gamma}{2}$ we can write the beam splitter transformation matrix from Eq.(3.30) as

$$\Lambda = \begin{pmatrix} \cos\frac{\gamma}{2} & \sin\frac{\gamma}{2} \\ \sin\frac{\gamma}{2} & -\cos\frac{\gamma}{2} \end{pmatrix} \tag{3.34}$$

which is almost equivalent to $R_y(\gamma)$. By inserting a phase shifter before the beam splitter we get the exact expression for $R_y(\gamma)$:

$$\begin{pmatrix} \cos\frac{\gamma}{2} & \sin\frac{\gamma}{2} \\ \sin\frac{\gamma}{2} & -\cos\frac{\gamma}{2} \end{pmatrix} \begin{pmatrix} 1 & 0 \\ 0 & -1 \end{pmatrix} = \begin{pmatrix} \cos\frac{\gamma}{2} & -\sin\frac{\gamma}{2} \\ \sin\frac{\gamma}{2} & \cos\frac{\gamma}{2} \end{pmatrix} = R_y(\gamma) \tag{3.35}$$

Thus, $R_y(\gamma)$ is implementable using a beam splitter and a phase shifter. To show that $R_z(\beta)$ is implementable we can rewrite Eq.(3.11) to

$$R_z(\beta) = e^{-i\frac{\beta}{2}} \begin{pmatrix} 1 & 0 \\ 0 & e^{i\beta} \end{pmatrix} \tag{3.36}$$

which can be realized with a phase shifter up to a global phase shift.

# Chapter 4

# Qubit representations

Qubits can be represented in many ways. One possibility is to use photons. This is a good choice because photons are chargeless particles which do not interact very strongly with either each other or most matter. Also, there is essentially no decoherence with photons. The only problem with photons is that controlled interactions between them are difficult to implement. However, by using nonlinear optics (optical media mediate interactions) you can make photons interact with each other rather easily. The problem is that the non-linear interaction is difficult to implement. Therefore one try to use linear optics, which work as passive devices, to cause interaction between modes (mainly by using beam splitters and phase shifters).
Here some of the most used encoding strategies will be mentioned, and the use of linear optics in some of these representations will be put in focus in Chapter 7.

## 4.1 Photon encodings

Encoding means that the same physical number state, depending on the encoding we choose, can be associated with different logical case.

### 4.1.1 Single rail

With single rail encoding strategy the qubit is encoded in the occupation, or non- occupation, of a single optical mode, similar to a classical computer hard drive where information are stored as a hole or not. This means that the vacuum state, $|0\rangle$, represents the logical zero while the single photon state, $|1\rangle$, represents the logical one. Several experiments have managed to create superpositions of states like this [RLS02, LM02].

This is a very compact representation with at most $n$ photons and only $n$ ports in a $n$-qubit gate. The problem with this representation is that different states have different photon-numbers (The logical state equals the number state), and with linear optics we cannot just create or destroy photons. But by using ancillary ports, this problem can be dealt with.

### 4.1.2 Dual rail

The largest difference between single rail and dual rail lie in the build-up of the dual rail. With dual rail the photon number is the same for all logical states. The logical 0 is represented by a single photon occupation of one mode with the other in the vacuum state. The logical 1 is the opposite of the logical 0 state with a single photon in the other mode. This means that the logical zero, $|0\rangle_L$ equals the state $|10\rangle$ (one photon in the first pulse and the vacuum state in the other) while the logical one, $|1\rangle_L$ equals the state $|01\rangle$.

In LOQC, dual rail logic is often implemented using the horizontal and vertical polarization modes of a single spatial mode. In polarization-encoded qubits the values of the qubits are represented by the two linear states of polarization of a single photon, with a horizontal polarization $|H\rangle$ representing the logical value 0, and a vertical polarization $|V\rangle$ representing the value 1.

If you use polarization modes, it is important to change the conventional beamsplitters to polarization beamsplitters as shown in Fig. 4.1.

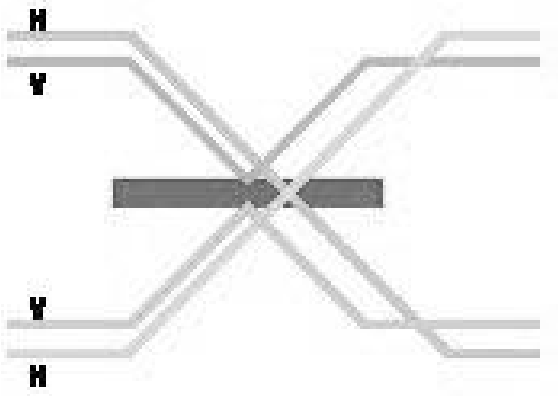

**Figure 4.1:** *A simple principle sketch of a polarization beamsplitter.*

Arbitrary unitary transforms can be applied to quantum information, encoded with single photons in the $c_0|01\rangle + c_1|10\rangle$ dual-rail representation, using phase shifters, beamsplitters and nonlinear optical Kerr media.

If we want to represent $n$ qubits with this representation we will need $N = 2n$ ports and $n$ photons. I.e. compared to single rail you have to double the number of ports.

Example of this representation:

| Logical state: | Number state: |
| --- | --- |
| $\|00\rangle_L \longrightarrow$ | $\|10\rangle\|10\rangle = \|1010\rangle$. |
| $\|11\rangle_L \longrightarrow$ | $\|01\rangle\|01\rangle = \|0101\rangle$. |

### 4.1.3 One hot

Taken to the extreme, you can devise an encoding scheme that has one bit per state. This type of encoding is called one-hot because only one state variable bit is set, or "hot," for each state. The benefit is that the next state generation function is simple. Also, because only two bits alter per transition, power consumption is small. The drawback with this encoding is the demand of register bits, one per state [Cyl00]. The equivalent encoding here, for a $m$-qubit state, would be a multiport with $2^m$ ports and just one photon in the corresponding port.

The problem with this representation is mainly the exponentially increase of multiports as a function of the number of qubits. But for just one or two qubits, one hot only need the same number of multiports as e.g. dual rail.

Example of this representation:

| Logical state: | Decimal number: | Number state: |
| --- | --- | --- |
| $\|00\rangle \longrightarrow$ | $0 \longrightarrow$ | $\|1000\rangle$ |
| $\|11\rangle \longrightarrow$ | $3 \longrightarrow$ | $\|0001\rangle$ |

# Chapter 5

# Accomplishments within LOQC

In this chapter we will take a closer look at the accomplishments in linear optics computing with today's technology, together with some of the problems that need to be solved in the future.

## 5.1 Advantages and obstacles

Here some of the advantages and problems in relation to linear optics computing will be listed:

| Advantages: | Obstacles: |
|---|---|
| Can be used at room temperature | Need for nonlinear interactions? |
| Easy to observe interference | Inefficient photoelectrons |
| Coherence time seems to be long | Inefficient one-photon sources |
| No lack of photons | Scalable mode matching |

There are several reasons for using linear optics instead of solid state quantum computers. With linear optics you can e.g. operate at room temperature, while solid state QC's so far need to operate at very low temperatures, to be in better control of the information carrying electrons. It is also easy to observe interference between photons, and so far it looks like the coherence time is long compared to the gate operation time. The fact that there never will be any lack of photons is another good property - but today's one-photon sources are rather inefficient.

One of the bigger problem is to get several entangled[1] photons without using nonlinear optics. In addition to this, mode matching is a huge problem. In nonclassical interference experiments, it is in general difficult to mode match, and this may be identified as a major contribution to nonunit visibility [RLBW02].

## 5.2 Can LOCQ satisfy the DiVincenzo Criteria today?

In Chapter 1.3 we listed the seven DiVincenzo criteria which show the *general* implementation-requirements of quantum computation. Now we consider the criteria in essence of LOQC . We notice that in criteria 1,3 and 7 a potentially viable approach has achieved sufficient proof of principle. The other criteria lack a sufficient proof of principle, even if a viable approach has been proposed. We will take a more thoroughly look at this below.

1. **Scalable physical system with well characterized qubits:**

   Qubits are usually represented by two different modes $|0\rangle$ and $|1\rangle$. This can among others be accomplished by the use of polarization modes.

2. **The ability to initialize the state of the qubits to a simple fiducial state:**

   The problem here is to get fast, reliable and "on-demand" single photon sources.

3. **Long relevant decoherence times, much longer than the gate operation time:**

   The decoherence-sources or sources of error do among others come from interferometric stability, mode matching, photon loss and detector accuracy and efficiency. By making more effective single photon sources and very effective photon detectors the decoherence time will be increased. However, the time it takes light to pass trough an optical element for single-qubit gates is very short, i.e. less than a picosecond. This is more of a problem for two-qubit gates and the improvement of this technology is a challenge.

[1]Entanglement: The property of two or more quantum systems whose total quantum state cannot be written as a product of the states of the individual systems (c.f., separable state); this property introduces nonlocality into quantum theory, and is believed to be an essential ingredient of quantum information processing. Mathematically it can be written $|\Psi\rangle = \frac{|00\rangle+|11\rangle}{\sqrt{2}}$ Here represent the value to the left of $|00\rangle$ and $|11\rangle$ one part which goes in one direction and the right goes in the other direction- in this way, they are "entangled."

4. **A universal set of quantum gates:**

   By using beam splitters, polarization rotators and phase shifters you can perform the most single-qubit operations. However, with two-qubit interactions you start to get problems because this is induced by measurement of photons in LOQC. This require an efficient photon counter, which is yet to be demonstrated. Also teleportation gates with some form of photon storage might be needed.

   Another problem is to produce entanglement on demand. By the use of nonlinear optics entangled states can be produced, but this doesn't happen on demand. This is important for implementing particular error-correction codes.

5. **A qubit-specific measurement capability:**

   To be able to implement fault-tolerance of LOQC, high efficiency(>99%), discriminating, single-photon devices are needed.

6. **The ability to interconvert stationary flying qubits:**

   Via the use of electro-optic devices, e.g. exciton quantum dots, optical schemes can interface to solid-state systems, but till now, no such detailed scheme has been demonstrated.

7. **The ability to faithfully transmit flying qubits between specified locations:**

   This is relative straightforward for free-space propagation of photons, but at the beam splitters it is necessary to have good mode matching. You can accomplish this by having well-defined mode structures.

## 5.3 Experimental accomplishments till today

Below we have listed some of the most important accomplishments in LOQC so far.

1. **Preparation and readout of qubit states**

   - By using conditional single-photon states from spontaneous parametric down conversion (SPDC) and postselection this requirements have been accomplished to some extent. More recently promising results from single quantum dots have been reported.

However the lowest achieved error (probability of somehting other than one photon) is still 60%.

- Single photon detectors with an efficiency of 88% have been demonstrated.
- Suggestions for greater than 99%-efficient detectors with photon-number resolving capability have been proposed [Ima02]. This is needed for a fault-tolerant implementation.

2. **Single-qubit operations:**

   - Rabi flops of a qubit have been demonstrated by a single-photon gate by the use of beam splitters.
   - Decoherence times much longer than Rabi oscillation[2] have also been demonstrated for a single-photon. (A huge number of single-qubit rotations could be done before photon loss became a problem.)
   - Control of both degrees of freedom on the Bloch sphere[MMRA02].

3. **Two-qubit operations:**

   - So far, one has partially achieved to make a nonuniversal two-qubit gate based on SPDC and postselection, but this is far from the desire to implement coherent two-qubit quantum logic operations.
   - With the use of SPCD and linear optics together with postselection one has achieved to produce and charaterize Bell states.
   - In high-finesse optical cavities decoherence time longer than the two-qubit time can be achieved.

4. **Operations on 3 to 10 physical qubits:**

   - A Greenberg, Horne and Zeilinger (GHZ) state[3] of three physical qubits have been achieved with the use of SPDC, linear optics and postselection.

[2]A two-state system driven by an electromagnetic wave whose energy equals the energy difference between the two states.

[3]This is an entangled states for a three-qubit system. There are two orthogonal GHZ states (with the form $|000\rangle \pm |111\rangle$)

- With the use of optical qubits, one has achieved to implement some simple quantum algorithms, such as Deutsch-Jesza [Tak00] and Grover [BvdHS02].

5. **Operations on one logical qubit:**

   - Not much have been achieved here. One of the problems is to keep a single logical qubit "alive" using repetitive error correction.

6. **Operations on two logical qubits:**

   - Not much have been achieved here either. What experimental physicists try to do are among others to produce two-logical-qubit Bell states and to implement two-logical-qubit operations.

7. **Operations on 3 to 10 logical qubits:**

   - For now, this also represent a huge problem. They do among others try to produce a GHZ state of three logical qubits, make a entangled state of four or more logical qubits, demonstrate the transfer of quantum information between logical qubits etc.

# Chapter 6

# Optical methods

In this chapter we will have a brief review of some different optical methods which may be important to realize LOQC. There are several of them, but there is just within a few of them that we've come far with today's technology.

## 6.1 Optical systems

### 6.1.1 Postselection

We can start with Knills definition of postselection [Kni01]:

G is a postselected gate with failure probability $f$ if each time G is used, it succeeds with probability $1 - f$ and fails otherwise, and whether it succeeded is known.

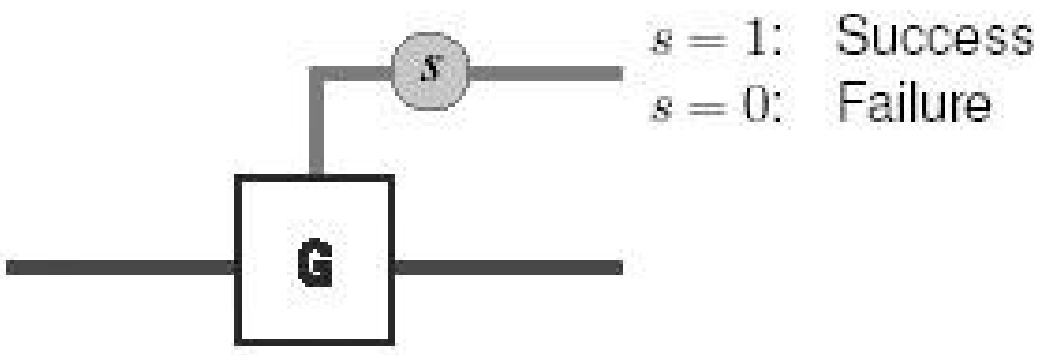

**Figure 6.1:** *A simple postselection gate.*

Whether or not it succeeds is known by looking at ancilla photons[1] in the gate. If the outcome of these ancillas is what you wish, the gate succeeds. In Chapter 7.1 we introduce the general formalism of postselection.

### 6.1.2 Postcorrection

Postcorrection or feedback is pre-determined combinations of phase corrections and bit-flips that are applied postselected output modes of non-deterministic quantum logic devices. In Chapter 8 we will introduce the general formalism of postcorrection before we implement this into a postselected gate.

### 6.1.3 Quantum teleportation

Quantum teleportation [BBC$^+$93] is a technique for moving quantum states around. Pursuant to the KLM article [KLM01], the success of quantum gates can be increased arbitrarily close to one with the use of non-deterministically prepared entangled states and quantum teleportation. Since quantum teleportation reduces to a state generation problem this technique might become important in LOQC.

A basic quantum teleportation protocol $T_1$ used to transfer a state $\alpha|0\rangle_1 + \beta|1\rangle_1$ from mode 1 to mode 3 begins by adjoining the ancilla states $|t_1\rangle_{23} = \gamma|01\rangle_{23} + \delta|10\rangle_{23}$. Afterwards, the two modes 1 and 2 are measured in the Bell basis[2] This measurement consist of two different steps. First a parity measurement which determines the parity $p$ of the numbers of photons in the two modes 1 and 2. Subsequently a sign check $s$ of the Bell state follows.

This give the following possible states in mode 3:

[1]Additional photons are sent through the logic operation at the same time as the state we are interested in. These extra photons are called ancilla photons and they trigger the optical circuitry that passes along the output of the logic operation when the result of the operation is correct. The ancilla photons are absorbed in photon detectors in the circuitry, but the output photons are preserved and passed on.

[2]For a quantum state with two subsystems (i.e. two qubits), the Bell states are the 4 orthogonal maximally entangled states (e.g., $|00\rangle + |11\rangle, |00\rangle - |11\rangle, |01\rangle + |10\rangle$ and $|10\rangle - |10\rangle$.

| Parity $p$ | Sign $s$ | State in mode 3 |
|---|---|---|
| Even | + | $\beta\|0\rangle_3 + \alpha\|0\rangle_3$ |
| Even | - | $-\beta\|0\rangle_3 + \alpha\|1\rangle_3$ |
| Odd | + | $\alpha\|0\rangle_3 + \beta\|1\rangle_3$ |
| Odd | - | $\alpha\|0\rangle_3 - \beta\|1\rangle_3$ |

A general quantum teleportation gate which in theory will work with a probability of 1 are shown in Fig. 6.2.

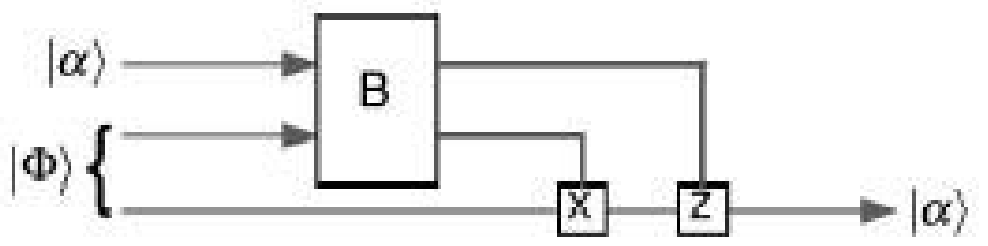

**Figure 6.2:** *A general quantum teleportation gate where an unknown state $|\alpha\rangle$ is teleported by making a joint Bell measurement B of this unknown state and one half of a Bell pair $\Phi$.*

Depending on the Bell measurement, a flip and/or phase shift can manipulate the other half of the Bell-pair $|\Phi\rangle$ and the unknown state is reconstructed. However, this transformation isn't to easy to implement only by using linear optics. The possible outcome with odd parity and a negative sign can easily be restored to the initial state by just using a phase shifter. The problems come when measuring even parity (where $|0\rangle$ and $|1\rangle$ are flipped). A state can not easily be un-flipped back to the initial state only by using linear optics.

By applying a balanced beamsplitter to both mode 1 and 2 at the same time as you measure the number of photons in the two modes, you can implement this teleportation protocol in linear optics with a probability of 1/2. By using this method it is also possible to implement a conditional sign flip with probability of 1/4. A setup for this is shown in Fig. 6.3 below.

Here we got two incoming qubits $Q_1$ and $Q_2$. The first mode of each qubit (1 and 3) is then being teleported to two modes (6 and 8) before the sign shift is applied to this new modes. Since each teleportation succeed with a probability of 1/2 the probability of two successful teleportations is 1/4. In addition to this probability, we also need to prepare the states and this needs on average to be attempted 16 times before success, which corresponds to 32 attempted NS operations.

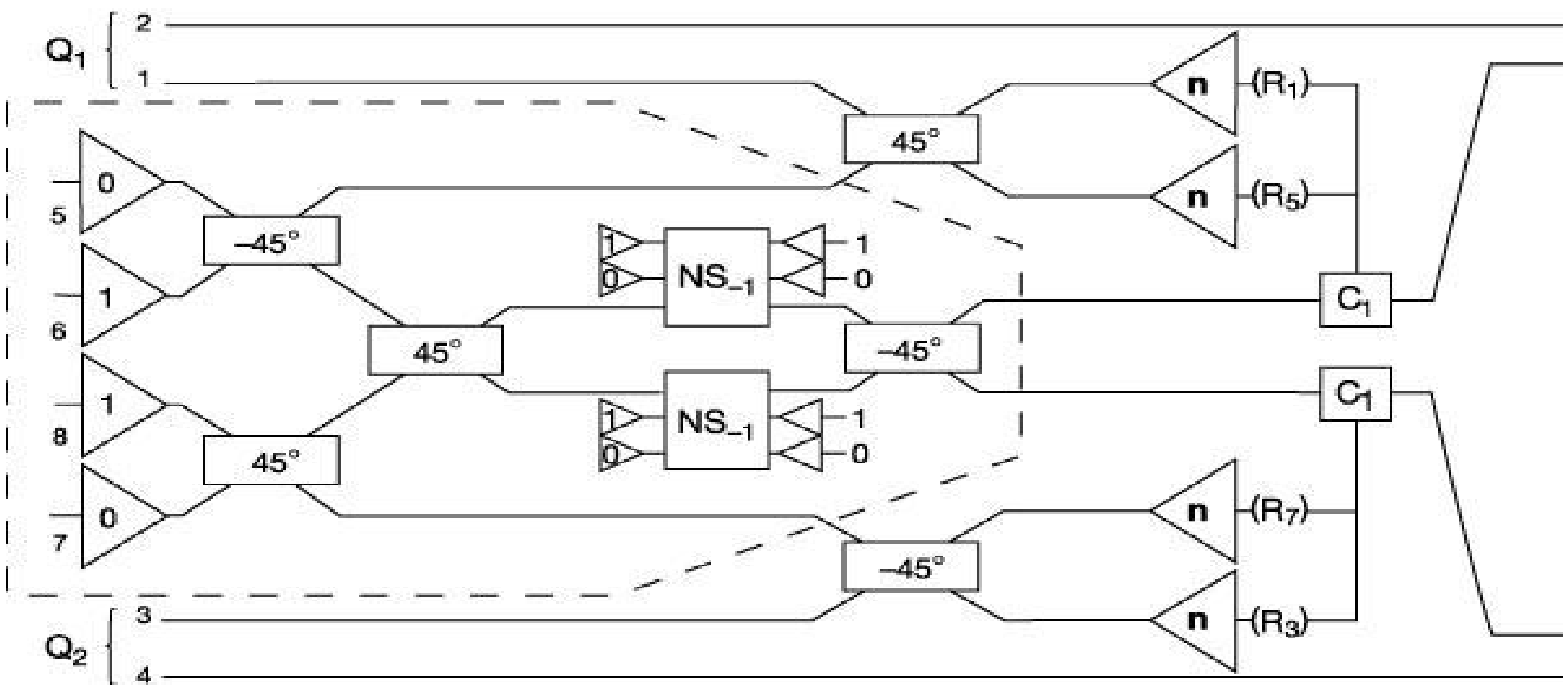

**Figure 6.3:** *A quantum gate giving a conditional sign flip with success probability of* $1/4$ *The operator* $C_1$ *is a* $180°$ *phase shifter which is added if the parity is odd and the sign is negative.* $NS_{-1}$ *just represent an ordinary* $NS(180°)$ *gate [KLM01].*

### 6.1.4 Quantum Error Correction and accuracy threshold

It is very important to protect quantum information against different types of errors, e.g. decoherence, calibration errors and random fluctuations in the control parameters. The idea is now to construct quantum error-correcting codes [Sho95, Ste96a], that will make it possible to quantum-compute in a fault-tolerant manner because error correction process itself induces an error.

In a simplified manner we can say that an error, like any physical process, is a unitary transformation of the state space. The space of errors to a single qubit is 4-dimensional, and is spanned by the four unitary matrices

| | |
|---|---|
| I | No error |
| X | Bit error |
| Z | Phase error |
| Y=iXZ | Combination |

where the Pauli matrices $X$,$Y$ and $Z$ (from Eq. 3.6) and $I$ now become noise operators.

Quantum error-correcting codes are defined as subspaces of the qubits' state space with the property that an initial state in this subspace can be recovered if sufficiently few of the qubits have experienced errors. Provided the noise affecting different qubits is independent and not too intense, any state stored in the subspace can the be regained with high fidelity.

It have been shown that if the effects of all errors per qubit and step of the computation are below a certain threshold, it is possible to process quantum information arbitrary accurately with reasonable resource overheads [ABO99, CRSS97, Got96].

Recently, T.Ralph proposed a quantum error correction gate in linear optics [Ral03]. The gate requires three CNOT-gates, and by using the non-deterministic CNOT proposed by Knill *et al.* [KLM01], the overall success rate of this gate would be 1/4000. This shows that the technical requirements for producing a quantum error correction gate with linear optics is way to high for today's technology.

# Chapter 7

# Postselection

One of the main problems of today is that even "ideal" linear optics quantum gates are non-deterministic, that is the probability of success is too low to get a functional benefit. In Chapter 6 we briefly introduced postselection and postcorrection as fruitful techniques for increasing the probability of gate-success. In this chapter we will introduce a general theory of postselection before we show the realizations of two important postselected gates.

## 7.1 General formalism of postselection

A postselection device can in general be described as a optical multiport composite system with two distinct physical systems,$\mathcal{H}_1$ and $\mathcal{H}_2$, where $\mathcal{H}_1$ is the state space of the input computational states and $\mathcal{H}_2$ is the state space of the input ancilla states, where the tensor product constitute the Hilbert space

$$\mathcal{H} = \mathcal{H}_1 \otimes \mathcal{H}_2. \tag{7.1}$$

as shown in Fig. 7.1.

When the projective measurement of interest is done, after a unitary evolution $U$, we can factorize the Hilbert space into a new output computing space $\mathcal{H}_{\bar{1}}$ and a new ancilla space $\mathcal{H}_{\bar{2}}$,

$$\mathcal{H} = \mathcal{H}_{\bar{1}} \otimes \mathcal{H}_{\bar{2}}. \tag{7.2}$$

The projective measurements are described by the projector

$$\bar{P} = I_{\bar{1}} \otimes \sum_{\bar{k}} s_{\bar{k}} |\bar{k}\rangle\langle\bar{k}| \tag{7.3}$$

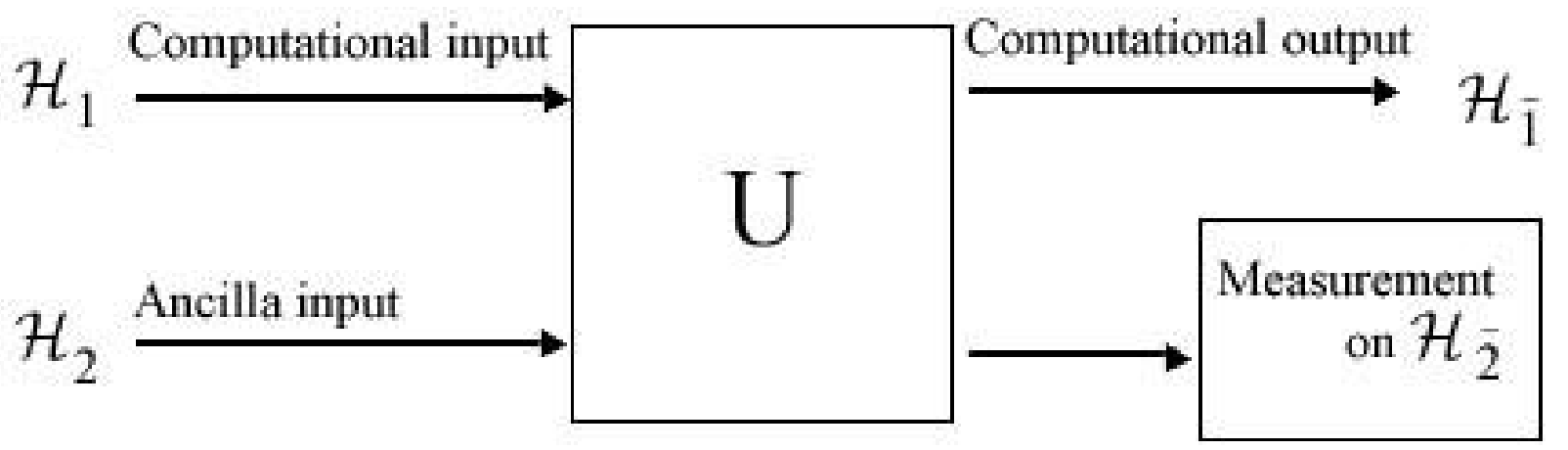

**Figure 7.1:** *A schematic of a post-selection device. The input computational channels, $\mathcal{H}_1$, and input ancilla channels, $\mathcal{H}_2$, undergo unitary evolution. The measurement performed in the ancilla output state space, $\mathcal{H}_{\bar{2}}$ indicates the success or failure of computation.*

where $\bar{I}$ is the identity operator in $\mathcal{H}_{\bar{1}}$, $|k\rangle$ is a orthonormal basis spanning the Hilbert space $\mathcal{H}_{\bar{2}}$ and $s_{\bar{k}}$ is equal to zero or unity.

The total density operator of the input can be written as

$$\rho^{12} = \rho \otimes \sigma \tag{7.4}$$

with $\rho$ as the initial density operator for the computational channels, and $\sigma$ the initial density operator for the ancilla channels, assuming that the ancillas are not entangled with the computational qubits. At the end of the unitary evolution the density operator is given by $U(\rho \otimes \sigma)U^\dagger$.
The probability of success depends on the projector $\bar{P}$

$$d(\rho) = tr_{\bar{1},\bar{2}}(U(\rho \otimes \sigma)U^\dagger \bar{P}). \tag{7.5}$$

and in the event of a successful measurement, the output of the channels in the $\bar{\mathcal{H}}_1$ state space can be described by the reduced density operator

$$\bar{\rho} = \frac{tr_{\bar{2}}(\bar{P}U(\rho \otimes \sigma)U^\dagger \bar{P})}{tr_{\bar{1},\bar{2}}(U(\rho \otimes \sigma)U^\dagger \bar{P})}. \tag{7.6}$$

We will not make use of this formalism directly, just state that postselection and postcorrection (Chapter 8.1) can be described in a compact general way. In [LKDS03] there exists calculation-examples where this formalism is applied directly.

### 7.1.1 The postselected non-linear sign shift gate

We will start by looking at the postselected non-linear sign shift, NS, gate [KLM01] because this might constitute the foundation of a CNOT gate.

Both the NS-gate and the two-mode conditional sign shift, CS, are examples on realizations of simple non-linear operations made possible just by the use of ancilla photons, linear optics and postselection.

When successful the NS implements the following transformation on a signal state $|\psi\rangle$:

$$NS(\theta) : \alpha|0\rangle + \beta|1\rangle + \gamma|2\rangle \rightarrow \alpha|0\rangle + \beta|1\rangle + e^{-i\alpha}\gamma|2\rangle \tag{7.7}$$

with $\alpha = 180°$.

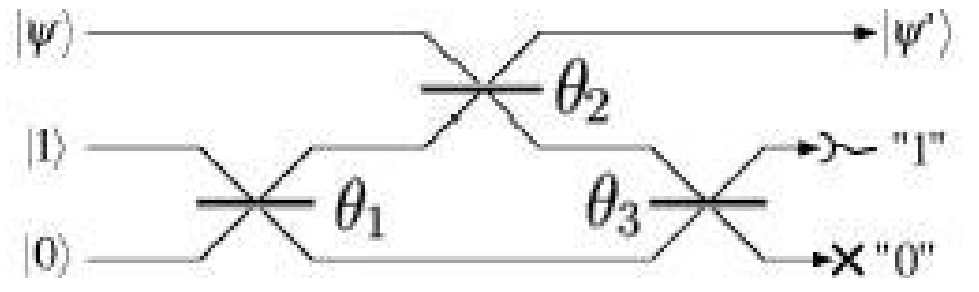

**Figure 7.2:** *An NS(*180°*). In the following example we use* $\theta_1 = \theta_3 = 22.5°$ *and* $\theta_2 = 65.53°$.

The pre-measurement evolution, which is done via the three beams splitters with pre-defined angles as in Fig. 7.2, is given by the unitary transformation $U$ and characterized by the transformation matrix

$$\Lambda = \begin{pmatrix} 1-\sqrt{2} & 1/\sqrt[4]{2} & \sqrt{3/\sqrt{2}-2} \\ 1/\sqrt[4]{2} & 1/2 & 1/2-1/\sqrt{2} \\ \sqrt{3/\sqrt{2}-2} & 1/2-1/\sqrt{2} & \sqrt{2}-1/2 \end{pmatrix}. \tag{7.8}$$

which can be calculated by applying the beam splitter relations in Eq.(3.31). Row one in the matrix represent the probability amplitudes where a photon comes out when it comes in to mode 1, column one equals the probability amplitude for a photon coming out from mode 1 and so on.

As can be seen from Fig. 7.2, the measure modes 2 and 3 only accept 1 and 0 photons, the incoming $|\psi\rangle$ in mode 1 may contain a superposition of 0,1 and 2 photons, and the incoming ancilla mode 2 contains one photon. We can now calculate the probability of a successful transformation of Eq.(7.7):

$$a_2^\dagger a_1^{\dagger k}|0\rangle_{321} \rightarrow \left[(1-\sqrt{2})a_1^\dagger + (1/\sqrt[4]{2})a_2^\dagger + (\sqrt{3/\sqrt{2}-2a_3^\dagger})\right]^k$$

$$\cdot \left[(1/2-1/\sqrt{2})a_3^\dagger + (1/2)a_2^\dagger + (1/\sqrt[4]{2})a_1^\dagger\right]|0\rangle_{321} \xrightarrow{\text{Mode 2,3 measure 1,0}} (\pm 1/2)a^{\dagger k}|0\rangle_1 \tag{7.9}$$

Here $k$ is either $0, 1$ or $2$ which represent the number of photons coming into mode 1 and the probability amplitude is $+1/2, +1/2, -1/2$ for $k = 0, 1, 2$ respectively.

The probability of success $_1\langle 0|(1/2)^2|0\rangle_1 = 1/4$ i.e. a 25% chance of success.

The two-mode conditional sign-shift gate, shown in Fig. 7.3 has the following transformation when successful:

$$CS(\theta) : \alpha|00\rangle + \beta|10\rangle + \gamma|01\rangle + \delta|11\rangle \rightarrow \alpha|00\rangle + \beta|10\rangle + \gamma|01\rangle + e^{-i\alpha}\delta|11\rangle \quad (7.10)$$

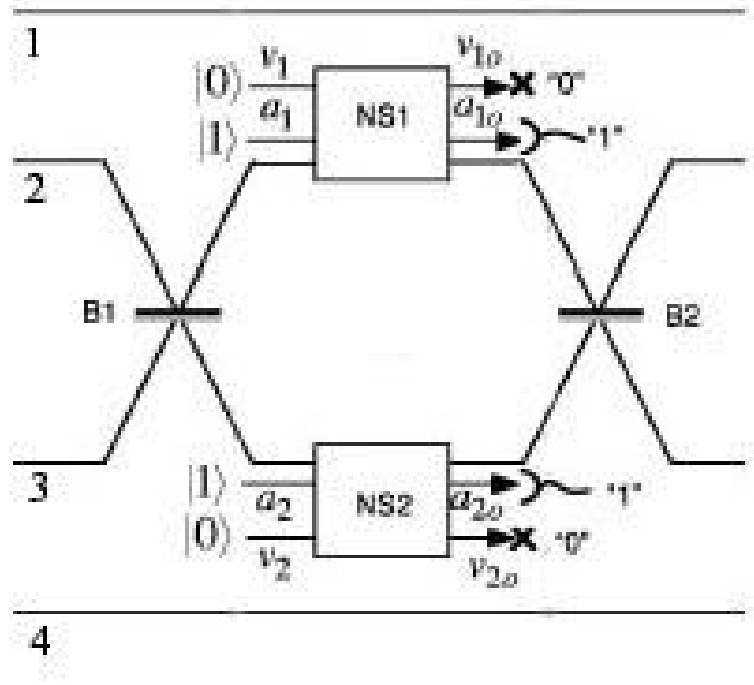

**Figure 7.3:** *A CS(*180°*) made by two NS-gates.*

If a $|0101\rangle_{4321}$ state is sent in, which corresponds to a logical $|11\rangle_L$, the probability of applying the phase shift successfully is 1/16, because this CS-gate is made of two NS-gates which both have to succeed and each got a 1/4 probability of success.

However, by the use of a so-called *efficient conditional sign flip* it is possible to get a probability of 2/27. This is among others shown by Knill [Kni02].

# Chapter 8

# Postcorrection

Having treated the general formalism of postselection in Chapter 7.1 , and having treated the NS - and CS gates in 7.1.1, we now naturally proceed towards a generalization of postcorrection before we propose a postcorrection scheme which hopefully will increase the probability of success of the basic NS-gate.

## 8.1 Generalization to include postcorrection

Sometimes we don't get the outcome of the ancilla we are supposed to. To correct this we can apply a deterministic unitary transformation on the output by sending different "adjustifying" classical messages depending on the outcome of the measurements. This is called postcorrection, shown in Fig. 8.1. This principle is also used in quantum teleportation.

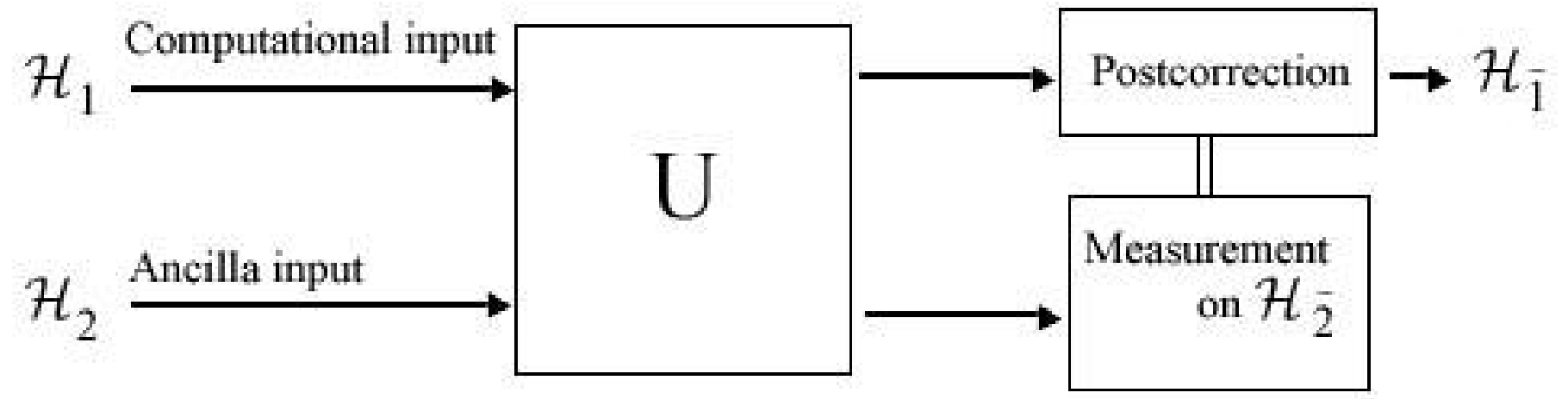

**Figure 8.1:** *A schematic of a postselection device that incorporates postcorrection. The double line represents classical channel that carries the measurement results. Based on the measurement outcome, the appropriate processing is performed on the output state space* $\mathcal{H}_{\bar{1}}$.

We now assume that the projective measurement is described by a set of projectors, each identifying a different detection signature, $\{\bar{P}_{(1)}, \bar{P}_{(2)}, ...\bar{P}_{(N)}, \bar{P}_{\perp}\}$, where

$$\bar{P}_{\perp} = I - \sum_{L=1}^{N} \bar{P}_{(L)} \tag{8.1}$$

and

$$\bar{P}_{L} = I \otimes \sum_{\bar{k}} s_{L,\bar{k}} |\bar{k}\rangle\langle\bar{k}|. \tag{8.2}$$

where $s_{L,\bar{k}}$ are equal to zero or unity such that

$$\bar{P}_{(L)}\bar{P}_{(L')} = \bar{P}_{(L)}\delta_{LL'} \tag{8.3}$$

We now get success if the projective measurement outcome is associated with *any* of the operators $\bar{P}_{(L)}$.
If outcome $L$ is achieved then the computational output is processed by application of the unitary operator $\bar{V}_{(L)}$ acting over $H_{\bar{1}}$. The probability getting the outcome $L$ is

$$d_{(L)}(\rho) = tr_{\bar{1},\bar{2}}(U(\rho \otimes \sigma)U^{\dagger}\bar{P}_{(L)}). \tag{8.4}$$

If outcome $L$ is achieved the postcorrected computational output is

$$\bar{\rho}_{(L)} = \frac{\bar{V}_{(L)}[tr_{\bar{2}}(\bar{P}_{(L)}U(\rho \otimes \sigma)U^{\dagger}\bar{P}_{(L)})]\bar{V}^{\dagger}_{(L)}}{tr_{\bar{1},\bar{2}}(U(\rho \otimes \sigma)U^{\dagger}\bar{P}_{(L)})}. \tag{8.5}$$

The conditions for these multiport devices to simulate unitary evolution have been derived in [LKDS03] .One of the conditions is that the probability of each successful outcome must be independent of the input density operator $\rho$. If the probability for success $d(\rho)$ of the measurement were dependent of the input density operator, by monitoring the success rate in an assembly of experiments all characterized by the same input $\rho$, one could learn something about $\rho$, and we would not expect operationally unitary evolution in the presence of this kind of gain of information.

## 8.2 Postcorrection in NS-gates

We can start by looking at the NS-gate discussed in Chapter 7.1.1, and modify this gradually. The basic NS doesn't accept any photons in mode 3. In our new suggested scheme the photon counter in mode 3 can accept 0 or 1 photon. In spite of this we can still use the same transformation matrix as in Eq.(7.8).

What we now need to do is to consider all the different possible outcomes and consider which output states we can correct.

If we detect two or more photons in our detectors, the incoming state $|\Psi\rangle$ can't be corrected. By measuring it we "know" it don't contain any $|0\rangle$, but what if it originally was a superposition state between e.g. $\alpha|0\rangle + \beta|1\rangle$? Then we will get $(\alpha + \beta)|0\rangle$ out, and it is impossible to correct this back to $\alpha|0\rangle + \beta|1\rangle$. Therefore we have to exclude all possibilities where we measure more than one photon in one or both of the detectors. We don't know the incoming state, but it is still on the form $\alpha|0\rangle + \beta|1\rangle + \gamma|2\rangle$. The general form of the remaining interesting outcomes are:

**Case 1:** Where we don't measure any photons in the detectors which correspond to $n_1 = n_2 = 0 : \alpha'|1\rangle + \beta'|2\rangle + \gamma'|3\rangle$

**Case 2:** Where we measure one photon in the middle detector, but no one in the lower which correspond to $n_1 = 1$ and $n_2 = 0$: $\alpha'|0\rangle+\beta'|1\rangle+\gamma'|2\rangle$

**Case 3:** Where we measure one photon in the lower detector, but no one in the middle which correspond to $n_1 = 0$ and $n_2 = 1$: $\alpha'|0\rangle+\beta'|1\rangle+\gamma'|2\rangle$

where $|\alpha'|^2 + |\beta'|^2 + |\gamma'|^2 \leq 1$.

When we now apply the unitary transformation matrix from Eq.(8.7) on the incoming state $|\psi\rangle = \alpha|0\rangle + \beta|1\rangle + \gamma|2\rangle$ we will end up with the following outgoing states[1] $|\Psi'\rangle$ (and for simplicity we have used $\alpha = \beta = \gamma = 1$ and ignored the normalization-factor in the further calculations):

**Case 1:** $\frac{1}{\sqrt[4]{2}} a_1^\dagger|0\rangle_{321} + (1-\sqrt{2}) \cdot \sqrt[4]{2} a_1^{\dagger^2}|0\rangle_{321} + \frac{(1-\sqrt{2})^2}{\sqrt[4]{2}} a_1^{\dagger^3}|0\rangle_{321}$
$\approx 0.84 a_1^\dagger|0\rangle_{321} - 0.70 a_1^{\dagger^2}|0\rangle_{321} + 0.25 a_1^{\dagger^3}|0\rangle_{321}$

**Case 2:** $\frac{1}{2} a_2^\dagger|0\rangle_{321} + \frac{1}{2} a_2^\dagger a_1^\dagger|0\rangle_{321} - \frac{1}{2} a_2^\dagger a_1^{\dagger^2}|0\rangle_{321}$

**Case 3:** $(\frac{1}{2} - \frac{1}{\sqrt{2}}) a_3^\dagger|0\rangle_{321} + \left[(1-\sqrt{2}) \cdot (\frac{1}{2} - \frac{1}{\sqrt{2}}) + \frac{1}{\sqrt[4]{2}} \cdot \sqrt{\frac{3}{\sqrt{2}} - 2}\right] a_3^\dagger a_1^\dagger|0\rangle_{321}$
$+ \left[\frac{1-\sqrt{2}}{\sqrt[4]{2}} \cdot \sqrt{\frac{3}{\sqrt{2}} - 2} + (1-\sqrt{2}) \cdot [(1-\sqrt{2})(\frac{1}{2} - \frac{1}{\sqrt{2}}) + \frac{1}{\sqrt[4]{2}} \cdot \sqrt{\frac{3}{\sqrt{2}} - 2}]\right] a_3^\dagger a_1^{\dagger^2}|0\rangle_{321}$
$\approx -0.21 a_3^\dagger|0\rangle_{321} + 0.38 a_3^\dagger a_1^\dagger|0\rangle_{321} - 0.28 a_3^\dagger a_1^{\dagger^2}|0\rangle_{321}$

[1]Maple calculations shown in Appendix A

In Case 2 what we want to happen comes strait out, as calculated earlier in Eq.(7.9), while we have to try to postcorrect the two other outcomes.

The postcorrection to Case 1 and Case 3 need to correct both the signs and the difference in probability amplitude within the superposition. We will try to accomplish this by applying different postcorrection-setups. First we can start with the simplest possible setup, just one beamsplitter.

### 8.2.1 Postcorrection with just one beamsplitter

In our fist attempt to apply postcorrection, we added a beamsplitter to a NS-gate, see Fig.8.2.

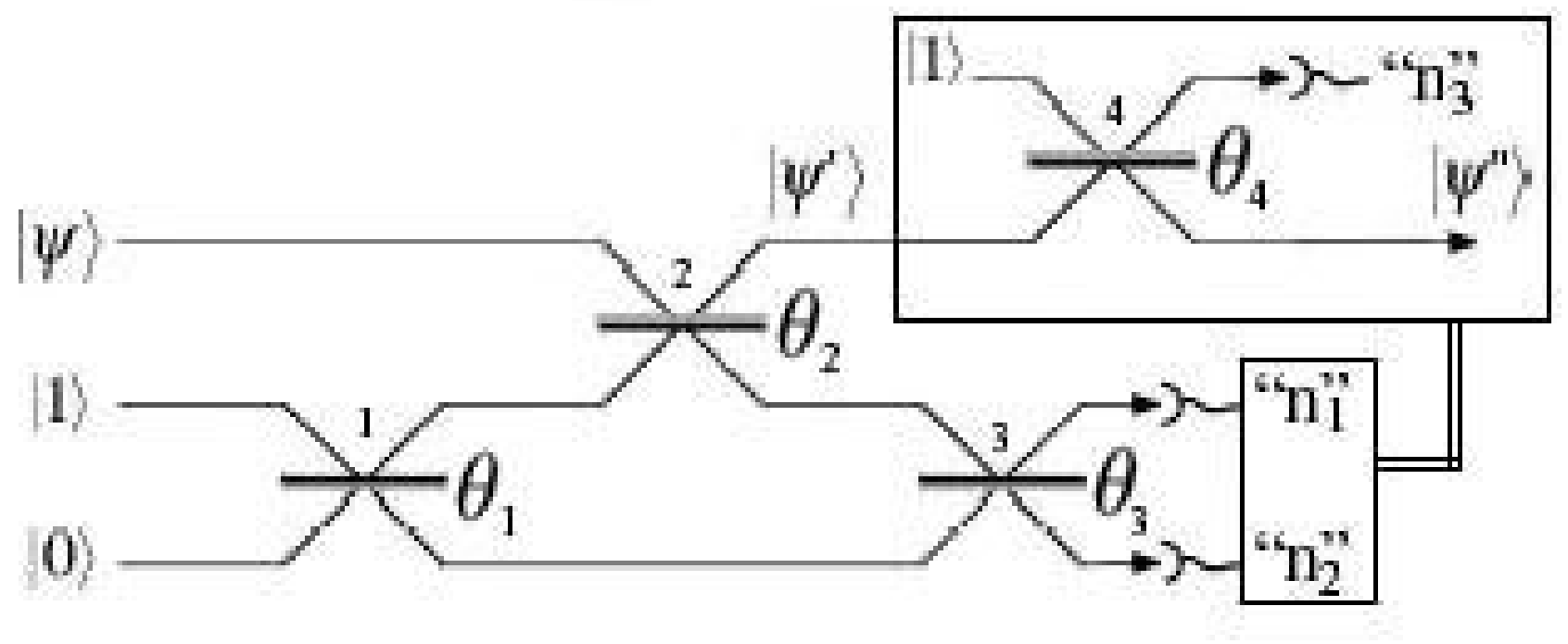

**Figure 8.2:** *An NS (*180°*), with just one beamsplitter trying to apply the postcorrection.*

With this setup we tried to apply two different types of postcorrection. First we tested if we could correct the state in the two cases to the form $c(\alpha|0\rangle + \beta|1\rangle - \gamma|2\rangle)$ where $c$ is the total probability for this transformation. When we found that this way of applying postcorrection failed[2], we tried to correct the states in the two cases back to the initial form, i.e. Case 1 and Case 3 back on the form $c(\alpha|0\rangle + \beta|1\rangle + \gamma|2\rangle)$ where $c$ is just a constant. If we manage to apply this transformation, we can just send the state to

[2]For the exact calculations with both our attempts to accomplish postcorrection with this setup, see Appendix B.

another NS-gate to increase the probability for success. However, also this attempt failed.

We can conclude with the fact that it is impossible to implement a correct postcorrection on this two cases with only one beamsplitter.

### 8.2.2 Postcorrection with two beamsplitters

Obviously we need to try something else; maybe we can manage something with two beamsplitters, one with the possibility to add an ancilla, and another beamsplitter which can remove additional photons. A phase shifter is also added, for the possibility to make sign-shifts this way, if it needs to be done. Fig. 8.3 shows the gate.

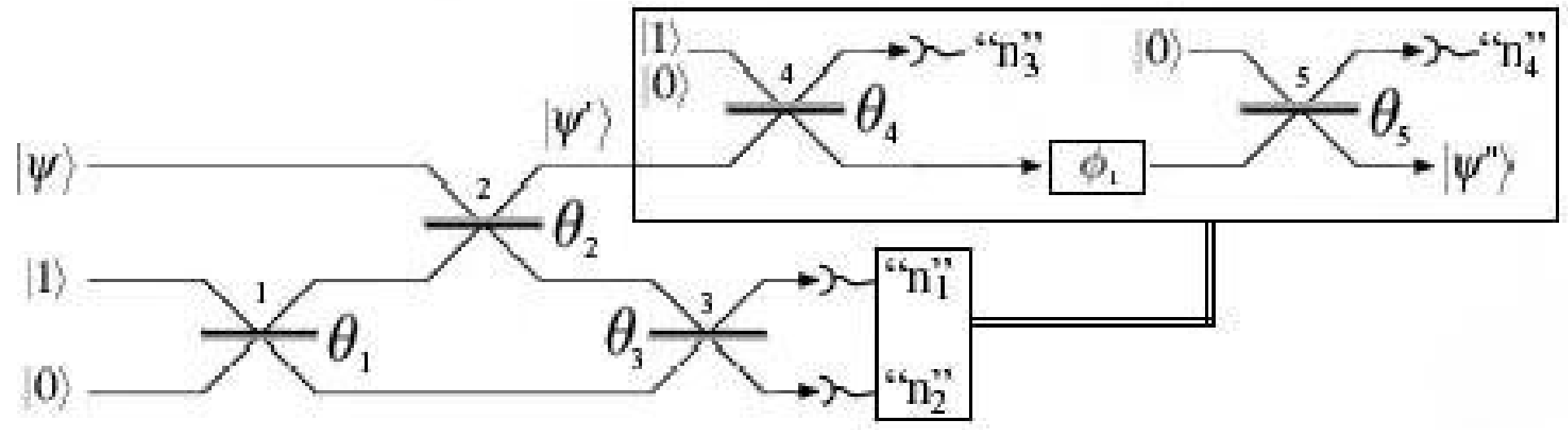

**Figure 8.3:** *An NS (180°) with the use of two beamsplitters doing the postcorrection.*

We tried to calculate this in Maple, but it didn't give out any answers so we're not sure if postcorrection can be implemented this way. What we need to solve in this calculation is e.g. for Case 1 the following problem:

We assume we got the original state $|0\rangle + |1\rangle + |2\rangle$ going trough the first NS-gate and we don't detect any photons. This will give $0.84|1\rangle_{321} - 0.70|2\rangle_{321} + 0.25|3\rangle_{321}$ coming out of the NS-gate (taken from Chapter 8.2). This need to be corrected in such a way that what we get out equals $c(|0\rangle + |1\rangle - |2\rangle)$, where $c$ is just a constant. For this to happen, the two beamsplitters have to operate on a general state $|0\rangle + |1\rangle + |2\rangle$ in such a way that we get $i|0\rangle + j|1\rangle + k|2\rangle$ after the postcorrection, where $|0.84 \cdot i| = |0.70 \cdot j| = |0.25 \cdot k| = c$. If this is possible, we might have a working postcorrection.

The only thing we discovered in our calculation was that the two new beamsplitters (4 and 5) need to have different $\theta$-values[3].

This means, we can not tell if it is possible to perform a correct postcorrection on this two cases with the use of two beamsplitters and possibly also a phase shifter.

### 8.2.3 Postcorrection applied in a new NS-gate

This is the last implementation of postcorrection we will have a look at. In this case we send the states in Case 1 and Case 3 to a new NS-gate to see if the postcorrection might be done there.

This give us three different beamsplitters in which we can change the $\theta$-values, and in the end hopefully this will give us a correct state-transformation.

If we e.g. consider Case 1 we got an additional incoming photon, which need to be removed simultaneously as we change the probability amplitudes. To reduce our possibilities, we assume we detect two photons in mode 2 in the new NS-gate[4].

By adding the probability amplitudes from the 2-photon outcome in mode 2 to the original state in Case 1, mentioned in Chapter 8.2, we get the following general out-coming state $|\psi''\rangle$:

$$
\begin{aligned}
|\psi''\rangle = \frac{1}{\sqrt[4]{2}} \cdot (\sin(t_1)\sin(t_2)\cos(t_3)\sin(t_3) + \cos(t_1)\cos(t_2)\sin(t_2)\cos(t_3)^2)|0\rangle \\
+(1-\sqrt{2})\cdot\sqrt[4]{2}\cdot[-2\sin(t_1)\cos(t_2)\sin(t_2)\cos(t_3)\sin(t_3) - 2\cos(t_1)\cos(t_2)^2\sin(t_2)\cos(t_3)^2 \\
+\cos(t_1)\sin(t_2)^3\cos(t_3)^2]|1\rangle + \frac{(1-\sqrt{2})^2}{\sqrt[4]{2}}\cdot \qquad (8.6) \\
[3\sin(t_1)\cos(t_2)^2\sin(t_2)\cos(t_3)\sin(t_3) + 3\cos(t_1)\cos(t_2)^3\sin(t_2)\cos(t_3)^2 \\
-3\cos(t_1)\cos(t_2)\sin(t_2)^3\cos(t_3)^2]|2\rangle
\end{aligned}
$$

We also tried to solve this equation, and the corresponding equation for Case 3, but we did not manage to get any interesting answers out of Maple here either[5].

Since there are a lot of possible solutions in this setup, we are quite sure it will be possible to get some interesting solutions which can give an implementation of a working postcorrection-setup. However, since we didn't manage to solve the equations we can't give any specific results. In e.g. Case 1, we might also get the right postcorrection when detecting two photons in mode 3 or one photon in mode 2 and one photon in mode 3 which leads to even more possible solutions.

[3] For our calculations, see Appendix C.

[4] The probability amplitudes for this operation are calculated in Appendix D.

[5] The calculations are shown in Appendix E.

## 8.3 Optimalization of postcorrection

We now proceed with a more general case in order to increase the success-probability even further. The general unitary transformation matrix can be expressed as follows:

$$\Lambda = \begin{pmatrix} -\cos(\theta_2) & \sin(\theta_2)\cos(\theta_3) & \sin(\theta_2)\sin(\theta_3) \\ \cos(\theta_1)\sin(\theta_2) & \sin(\theta_1)\sin(\theta_3)+\cos(\theta_1)\cos(\theta_2)\cos(\theta_3) & -\sin(\theta_1)\cos(\theta_3)+\cos(\theta_1)\cos(\theta_2)\sin(\theta_3) \\ \sin(\theta_1)\sin(\theta_2) & -\cos(\theta_1)\sin(\theta_3)+\sin(\theta_1)\cos(\theta_2)\cos(\theta_3) & \cos(\theta_1)\cos(\theta_3)+\sin(\theta_1)\cos(\theta_2)\sin(\theta_3) \end{pmatrix} \tag{8.7}$$

By calculating all the possible outcomes[6], we will get several terms which represent the total outgoing state space. However, only few of the outcomes are relevant and might be corrected. These outcomes corresponds to the same projective measurements as those described in Case 1,2 and 3 from Chapter 8.2.

Let us first consider the case where we transmit 0 photons into mode 1. Then we achieve a total of three possible outcomes. They are as follows:

$$\begin{gathered} -\cos(\theta_1)\sin(\theta_2)\mathbf{a_1^+}|\mathbf{0}\rangle \\ (\sin(\theta_1)\sin(\theta_3)+\cos(\theta_1)\cos(\theta_2)\cos(\theta_3))\mathbf{a_2^+}|\mathbf{0}\rangle \\ (\sin(\theta_1)\cos(\theta_3)-\cos(\theta_1)\cos(\theta_2)\sin(\theta_3))\mathbf{a_3^+}|\mathbf{0}\rangle \end{gathered} \tag{8.8}$$

Transmitting 1 incoming photon in mode 1 we also obtain three interesting outcomes:

$$\begin{gathered} \sqrt(2)\cos(\theta_1)\cos(\theta_2)\sin(\theta_3)\mathbf{a_1^+}^{\mathbf{2}}|\mathbf{0}\rangle \\ (\cos(\theta_1)\sin(\theta_2)^2\cos(\theta_3)-\sin(\theta_1)\cos(\theta_2)\sin(\theta_3)-\cos(\theta_1)\cos(\theta_2)^2\cos(\theta_3))\mathbf{a_2^+}\mathbf{a_1^+}|\mathbf{0}\rangle \\ (\cos(\theta_1)\cos(\theta_2)^2\sin(\theta_3)-\sin(\theta_1)\cos(\theta_2)\cos(\theta_3)-\cos(\theta_1)\sin(\theta_2)^2\sin(\theta_3))\mathbf{a_3^+}\mathbf{a_1^+}|\mathbf{0}\rangle \end{gathered} \tag{8.9}$$

And finally for 2 incoming photons in mode 1:

$$\begin{gathered} -\sqrt(3)\cos(\theta_1)\cos(\theta_2)^2\sin(\theta_2)\mathbf{a_1^+}^{\mathbf{3}}|\mathbf{0}\rangle \\ (\sin(\theta_1)\cos(\theta_2)^2\sin(\theta_3)+\cos(\theta_1)\cos(\theta_2)^3\cos(\theta_3)-2\cos(\theta_1)\cos(\theta_2)\sin(\theta_2)^2\cos(\theta_3))\mathbf{a_2^+}\mathbf{a_1^+}^{\mathbf{2}}|\mathbf{0}\rangle \\ (\sin(\theta_1)\cos(\theta_2)^2\cos(\theta_3)-\cos(\theta_1)\cos(\theta_2)^3\sin(\theta_3)+2\cos(\theta_1)\cos(\theta_2)\sin(\theta_2)^2\sin(\theta_3))\mathbf{a_3^+}\mathbf{a_1^+}^{\mathbf{2}}|\mathbf{0}\rangle \end{gathered} \tag{8.10}$$

[6]Exact calculation in Appendix F

By optimizing Eq.(8.8), Eq.(8.9) and Eq.(8.10), it is possible to find the angles which give the greatest probability of success.
Unfortunately, we didn't manage to optimize this numerically, and for this optimization to be interesting we also need to be able to implement some kind of postcorrection to it. So first it is necessary to get several possible optimal solutions. Afterwards we need to try to implement some kind of postcorrection to this solutions, and finally we can compare the results with the best probability so far, 1/4.

### 8.3.1 Postcorrection in CS-gates

As shown in Fig. 7.3 a CS-gate can be implemented by using two NS-gates. This means, if we succeed with postcorrection in the NS-gate, we can use this postcorrected NS-gate to construct a CS-gate. By doing this we get a CS-gate with postcorrection, at the same time as there will be an increase in the probability of success.

### 8.3.2 Postcorrection in CNOT

There are a number of proposals for implementing a non-deterministic CNOT-gate with linear optics and photodetetectors. The proposals require heralded entangled or deterministic single photon sources, and selective detectors that can distinguish with very high efficiency between zero, one and multiple photons.
The gate can e.g. be implemented by using the CS-gate, which means that it can be constructed as shown above, but with a few changes, as shown in Fig. 8.4.

Since the probability of success only depend on the CS-gate, the CNOT with postcorrection will work with a probability depending on the CS-gate discussed earlier. So if we manage to apply postcorrection in a NS-gate to increase the probability of success, this will also increase the probability of success in the CNOT-gate.

#### Upper bounds for NS and CS

We haven't managed to calculate any upper bounds for NS with postcorrection, but Knill have managed to find upper bounds for both NS- and CS-gates with use of postselection [Kni03]. It was shown that the upper bounds for a successful NS-gate was 1/2 and a CS-gate can possibly have a 3/4 probability of success. This also give a upper probability of 3/4 for a CNOT made by the use of a CS-gate, using only postselection.

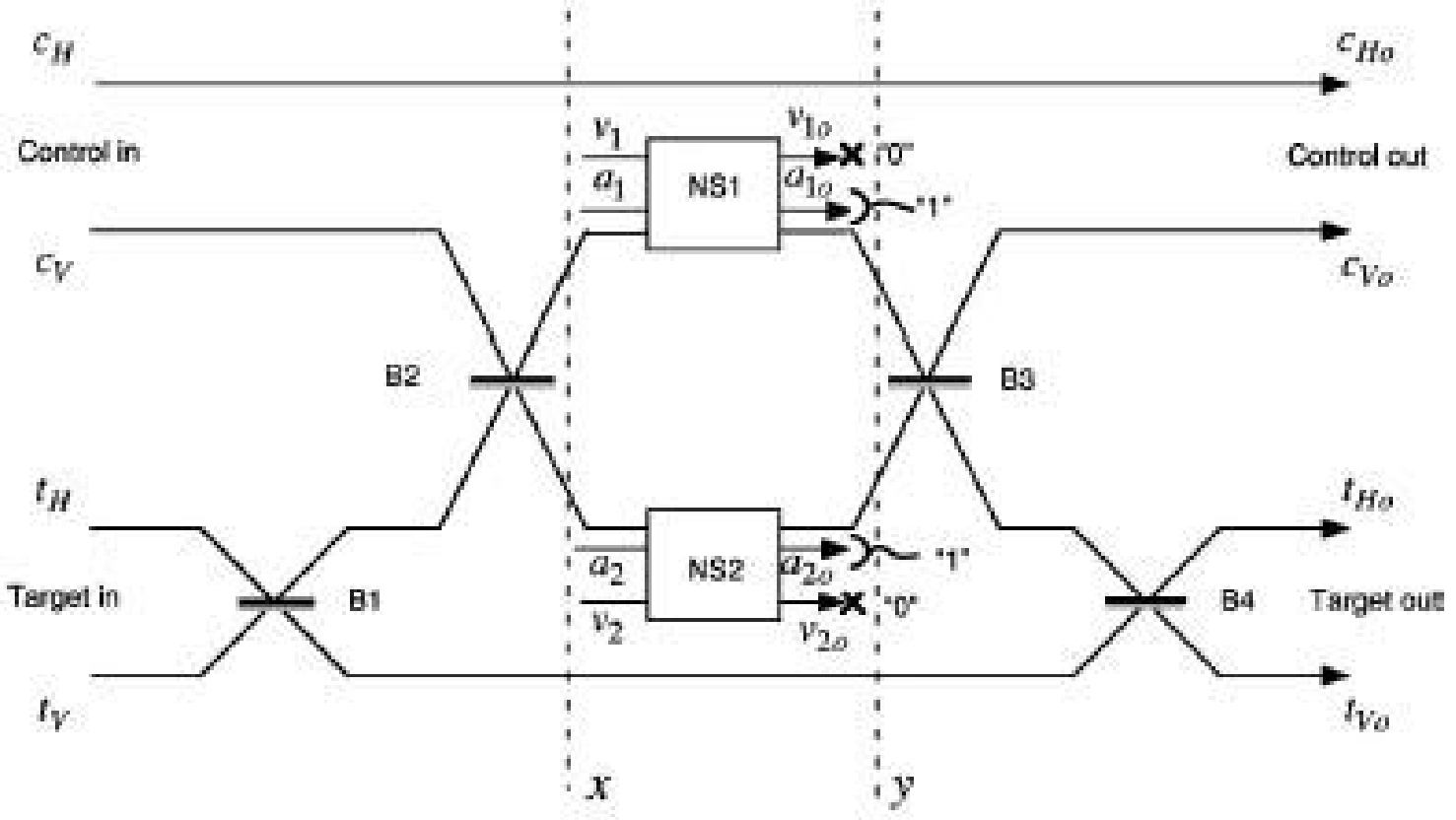

**Figure 8.4:** *A CNOT made by a CS-gate[KLM01].*

### 8.3.3 Polarization-encoded CNOT with two photons

In our earlier approach the qubit was represented by a single photon that was located in one of two paths, such as optical fibers. In the ideal case this is fine, but in reality it is known that the logic operations involve interference between different optical paths which can be very sensitive to thermally induced phase shift. The use of polarization-encoded qubits, mentioned in Chapter 4.1.2, can eliminate the need for any interference optical paths, which may be an advantage in practical applications

An interesting CNOT gate, with success probability $\frac{1}{9}$, that employ dual rail logic and only requires a two-photon source was presented in 2002 [RLBW02]. A gate similar to this has also been verified experimentally [RBPW02]. The gate also make use of postselection, and we find it appropriate to introduce it in contrast to the CNOT in Chapter 8.3.2. Will this gate be possible to postcorrect ?
We will explain the principles behind this gate, which is shown in Fig. 8.5.

The "control in" qubit is represented by the two bosonic operators $c_H$ and $c_V$. A single photon in $c_H$ with $c_V$ in a vacuum state will represent logical 0, which we will write $|H\rangle_c$, while a single photon occupation of $c_V$ with $c_H$ in a vacuum state will be the logical 1, which we denotes $|V\rangle_c$. The use of $H$ and $V$ to describe the states of the qubits alludes to the encoding in polarization. To go from polarization encoding to dual rail spatial encoding and vica versa in the lab requires a polarization beam splitter and half-wave plate.
The "target in" is represented by the mode operators $t_H$ and $t_V$ and the states $|H\rangle_t$ and $|V\rangle_t$. The ancillary vacuum input modes $v_c$ and $v_t$ complete the network.
Table 8.1 shows the swapping of the target modes which corresponds to the CNOT-operation $|x, y\rangle \rightarrow |x, x \oplus y\rangle$ where $\oplus$ is addition modulo two.

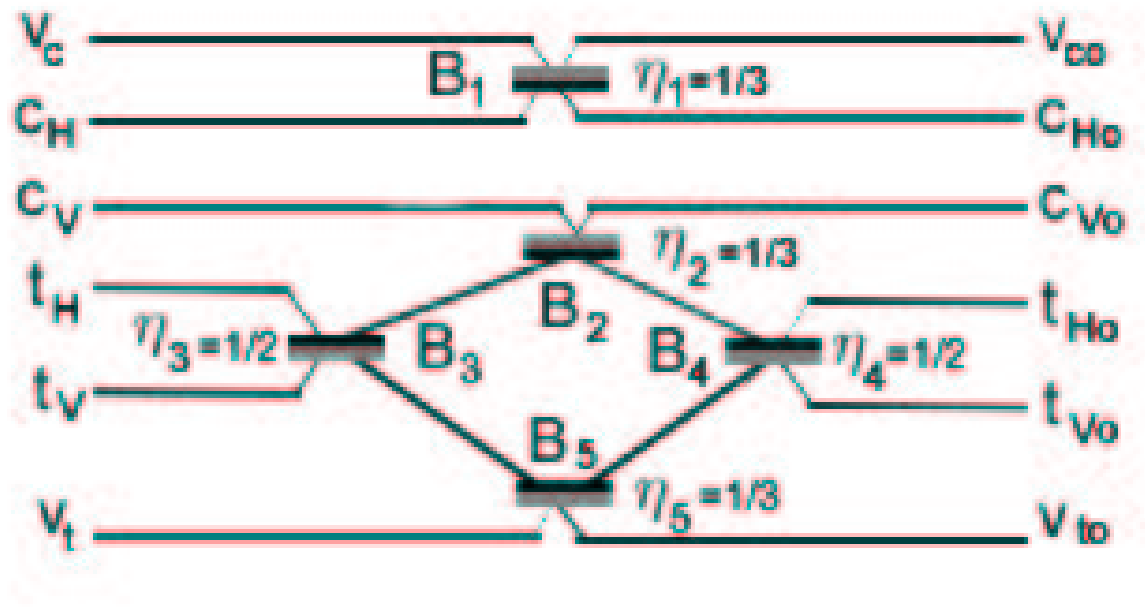

**Figure 8.5:** *A schematic of a non-deterministic two photon CNOT gate, where the beam splitter reflectivities $\eta$ and asymmetric phase shifts are indicated. The control modes are $c_H$ and $c_V$. The target modes are $t_H$ and $t_V$. The modes $v_c$ and $v_t$ are unoccupied ancillary modes.*

**Table 8.1:** *The intention of the gate is to swap the target modes if the control is in the state $|V\rangle_c$, but do not if the control is in the state $|H\rangle_c$.*

| In | Out |
|---|---|
| $\|H\rangle_c\|H\rangle_t$ | $\|H\rangle_c\|H\rangle_t$ |
| $\|H\rangle_c\|V\rangle_t$ | $\|H\rangle_c\|V\rangle_t$ |
| $\|V\rangle_c\|H\rangle_t$ | $\|V\rangle_c\|V\rangle_t$ |
| $\|V\rangle_c\|V\rangle_t$ | $\|V\rangle_c\|H\rangle_t$ |

With help from the Heisenberg equations as introduced in Eq.(3.31),

$$
\begin{aligned}
b_1 &= \sqrt{\eta}a_1 + \sqrt{1-\eta}a_2 \\
b_2 &= \sqrt{1-\eta}a_1 - \sqrt{\eta}a_2,
\end{aligned}
\tag{8.11}
$$

we can find the output modes that correspond to the input modes:

$$
\begin{aligned}
c_{H_o} &= \frac{1}{\sqrt{3}}(\sqrt{2}v_c + c_H), \\
c_{V_o} &= \frac{1}{\sqrt{3}}(-c_V + t_H + t_V), \\
t_{H_o} &= \frac{1}{\sqrt{3}}(c_V + t_H + v_t), \\
t_{V_o} &= \frac{1}{\sqrt{3}}(c_V + t_V - v_t), \\
v_{c_o} &= \frac{1}{\sqrt{3}}(-v_c + \sqrt{2}c_H), \\
v_{t_o} &= \frac{1}{\sqrt{3}}(t_H - t_V - v_t)
\end{aligned}
\tag{8.12}
$$

We can now consider a general input state

$$
\begin{aligned}
|\phi\rangle_{in} &= (\alpha|HH\rangle + \beta|HV\rangle + \gamma|VH\rangle + \delta|VV\rangle)|00\rangle \\
&= (\alpha c_H^\dagger t_H^\dagger + \beta c_H^\dagger t_V^\dagger + \gamma c_V^\dagger t_H^\dagger + \delta c_V^\dagger t_V^\dagger)|0000\rangle|00\rangle
\end{aligned}
\tag{8.13}
$$

where the ket-ordering is $|n_{c_H} n_{c_V} n_{t_H} n_{t_V}\rangle|n_{v_c} n_{v_t}\rangle$ and $n_{c_H} = c_H^\dagger c_H$ etc. and we use the use the shorthand $|1010\rangle = |HH\rangle, |1001\rangle = |HV\rangle$ etc.

The output-state can now be directly calculated by substituting the input-operators from Eq.(8.13) for the output operators given in Eq.(8.12),

$$
\begin{aligned}
|\psi\rangle_{out} &= (\alpha c_{H_o}^\dagger t_{H_o}^\dagger + \beta c_{H_o}^\dagger t_{V_o}^\dagger + \gamma c_{V_o}^\dagger t_{H_o}^\dagger + \delta c_{V_o}^\dagger t_{V_o}^\dagger)|0000\rangle|00\rangle \\
&= \frac{1}{3}[\alpha|HH\rangle|00\rangle + \beta|HV\rangle|00\rangle + \gamma|VV\rangle|00\rangle + \delta|VH\rangle|00\rangle + \sqrt{2}(\alpha+\beta)|0100\rangle|10\rangle \\
&+ \sqrt{2}(\alpha-\beta)|0000\rangle|11\rangle + (\alpha+\beta)|1100\rangle|00\rangle + (\alpha-\beta)|1000\rangle|01\rangle \\
&+ \sqrt{2}\alpha|0010\rangle|10\rangle + \sqrt{2}\beta|0001\rangle|10\rangle - \sqrt{2}(\gamma+\delta)|0200|00\rangle - (\gamma-\delta)|0100\rangle|01\rangle \\
&+ \sqrt{2}\gamma|0020\rangle|00\rangle + (\gamma-\delta)|0010\rangle|01\rangle + (\gamma+\delta)|0011\rangle|00\rangle + (\gamma-\delta)|0001\rangle|01\rangle \\
&+ \sqrt{2}\delta|0002\rangle|00\rangle]
\end{aligned}
\tag{8.14}
$$

From $|\phi\rangle_{out}$ we observe that the success probability for the CNOT-operation

$$\alpha|HH\rangle+\beta|HV\rangle+\gamma|VH\rangle+\delta|VV\rangle \xrightarrow{gate-transformation} \alpha|HH\rangle+\beta|HV\rangle+\gamma|VV\rangle+\delta|VH\rangle \tag{8.15}$$

, where the state $|\phi\rangle_{cb} = \alpha|HH\rangle + \beta|HV\rangle + \gamma|VV\rangle + \delta|VH\rangle$ is postselected in the coincidence basis, is $|\frac{1}{3}|^2 = \frac{1}{9}$
We have to consider this as an ideal probability where we have ignored errors in devices. Errors caused by timing of photons, non-optimal beam splitter ratios and mode matching that contributes to nonunit visibility will reduce the probability slightly further.
Unfortunately it seems unlikely that we can apply postcorrection on this scheme because we don't know which of the 9 possibilities we obtain after the postselection. We only now that five of nine state possibilities are unwanted, and reduces the probability of a successful CNOT operation. Considering a NS-gate we knew exactly which state we had to postcorrect after a successful postselection. This is not the case in this example. If we don't know what state to correct it is not easy to implement any form of postcorrection.

# Chapter 9

# Conclusions

## 9.1 Discussion

Although we are far from implementing a deterministic quantum gate, postselection and postcorrection represent important optical methods that should be further explored.

Even if we didn't manage to solve all the equations in relation to postcorrection there might be linear optics solutions that leads to an increase in the overall probability with two or more beam splitters in combination with phase shifters. However this setup must be implemented in a smart way. Since each beam splitter reduce the probability of a successful transformation, sending a classical message to several postcorrecting NS-gates in a row will only increase the probability by a small amount.

One of the most important questions today is whether or not it is possible to implement deterministic quantum gates by using postcorrection or other optical methods. So far there exist no proof that this neither can nor cannot be done deterministically, but we find it unlikely it can be done by only using postcorrection. Since it is impossible to correct the state if you detect more photons than incoming ancillas, the probability will naturally be reduced as a consequence of the coherence-loss.

The theoretical aspects are not the only problems behind the realization of scalable LOQC, i.e. where the capability of achieving the same efficiency is almost independent of the number of bosonic qubits. Without doubt there also exist many technological challenges before LOQC can be realized. E.g.

to implement the NS gate discussed in Chapter 7.1.1, progress is dependent on the availability of single-photon sources and discriminating single-photon detectors. The implementation of postcorrection also demands larger more complicated setups and this might be one of the reasons why this scheme is so unexplored.

## 9.2 Summary

An interesting reason to study quantum computing is that it is a new and insightful way to think about the fundamental laws of physics. The quest to bring together information, Turing machines and quantum mechanics is fascinating and what one could have the good fortune to encounter. It is already changing the way we understand and control matter at the atomic scale, making the fundamental quantum world more familiar and understandable.

In general we believe that quantum computing not will replace classical computing for similar reasons that quantum physics does not replace classical physics: No one takes their bicycle to be mended by a quantum mechanic. If larger quantum computers are made, they will be used to address just those special tasks which benefit from quantum information processing.

Since this project mainly was meant to be a literature study, the disposal time for new research was limited. However, we have shone light on some new postcorrection implementations, and hopefully contributed to bring some new thoughts to life.

Put together, it leads to the expectation that linear optical elements might play a fundamental role in future development of quantum communication and applications. The main obstacle to scalable optical quantum information processing lies in the requirement for nonlinear couplings between optical modes of few photons, a difficult experimental task.

# Appendix A

# Standard non-linear sign shift gate calculations

```
> with(linalg):

> a:=port1;
```

$a = port1$

$$a := port1$$

```
> b:=port2;
> c:=port3;
```

$$b := port2$$

$$c := port3$$

```
> u := matrix([ [1-sqrt(2), 1/(2^(1/4)),
> sqrt((3/sqrt(2))-2)],
> [ 1/(2^(1/4)), 1/2,1/2-1/(sqrt(2))],
> [ sqrt((3/sqrt(2))-2),1/2-1/(sqrt(2)),sqrt(2)-1/2]
> ]);
```

$$u := \begin{bmatrix} 1-\sqrt{2} & \frac{2^{(3/4)}}{2} & \frac{\sqrt{-8+6\sqrt{2}}}{2} \\ \frac{2^{(3/4)}}{2} & \frac{1}{2} & \frac{1}{2}-\frac{\sqrt{2}}{2} \\ \frac{\sqrt{-8+6\sqrt{2}}}{2} & \frac{1}{2}-\frac{\sqrt{2}}{2} & \sqrt{2}-\frac{1}{2} \end{bmatrix}$$

Definerer d,e og f som rad 1,2 og 3 i u-matrisen:

```
> d:=u[1,1]*a+u[1,2]*b+u[1,3]*c;
```

> rad 1

$$d := (1-\sqrt{2})\,\mathit{port1} + \frac{2^{(3/4)}\,\mathit{port2}}{2} + \frac{\sqrt{-8+6\sqrt{2}}\,\mathit{port3}}{2}$$

```
> e:=u[2,1]*a+u[2,2]*b+u[2,3]*c;
```

> rad 2

```
> f:=u[3,1]*a+u[3,2]*b+u[3,3]*c;
```

> rad 3

$$e := \frac{2^{(3/4)}\,\mathit{port1}}{2} + \frac{\mathit{port2}}{2} + (\frac{1}{2} - \frac{\sqrt{2}}{2})\,\mathit{port3}$$

$$f := \frac{\sqrt{-8+6\sqrt{2}}\,\mathit{port1}}{2} + (\frac{1}{2} - \frac{\sqrt{2}}{2})\,\mathit{port2} + (\sqrt{2} - \frac{1}{2})\,\mathit{port3}$$

## 0 fotoner i verste mode:

```
> evalf(expand(e));
```

$$0.8408964155\,\mathit{port1} + 0.5000000000\,\mathit{port2} - 0.2071067810\,\mathit{port3}$$

```
> total_sannsynlighet=.8408964155^2+.5000000000^2+.2071067810^2;
```

$$\mathit{total_sannsynlighet} = 1.000000000$$

## *Sannsynligheter:*

```
> .8408964155^2;
```

> 1 foton i port 1

```
> 0.5000000000^2;
```

> 1 foton i port 2

```
> 0.2071067810^2;
```

> 1 foton i port 3

```
> total_sannsynlighet=.8408964155^2+.5000000000^2+.2071067810^2;
```

$$0.7071067816$$

$$0.2500000000$$

$$0.04289321874$$

$$total_sannsynlighet = 1.000000000$$

## 1 fotoner i verste mode

```
> evalf(expand(d*e));
```

$$-0.3483106995\,port1^2 + 0.5000000000\,port1\,port2 + 0.378679656\,port1\,port3 + 0.4204482078\,port2^2 - 0.2\,10^{-9}\,port2\,port3 - 0.0721375076\,port3^2$$

Sannsynligheter:

```
> a[1]:=(sqrt(2))^2*0.3483106995^2;
```
> 2 fotoner i mode 1

$$a_1 := 0.2426406868$$

```
> a[2]:=.5000000000^2;
```
> 1 foton i mode 1 og 1 foton i mode 2

$$a_2 := 0.2500000000$$

```
> a[3]:=.378679656^2;
```
> 1 foton i mode 1 og 1 foton i mode 3

$$a_3 := 0.1433982819$$

```
> a[4]:=(sqrt(2))^2*.4204482078^2;
```
> 2 fotoner i mode 2

$$a_4 := 0.3535533908$$

```
> a[5]:=.2e-9^2;
```
> 1 foton i mode 2 og 1 foton i mode 3

$$a_5 := 0.4\,10^{-19}$$

```
> a[6]:=(sqrt(2))^2*.721375076e-1^2;
```
> 2 fotoner i mode 3

$$a_6 := 0.01040764001$$

```
> total_sannsynlighet2=sum('a[k]','k'=1..6);
```

$$total_sannsynlighet2 = 0.9999999995$$

2 fotoner i verste mode

```
> evalf(expand(d^2*e));
```

$-0.060660171\, \mathit{port3}^2\, \mathit{port2} - 0.5000000000\, \mathit{port1}^2\, \mathit{port2} - 0.2781745915\, \mathit{port1}^2\, \mathit{port3}$
$- 0.025126266\, \mathit{port3}^3 + 0.492585715\, \mathit{port3}\, \mathit{port1}\, \mathit{port2}$
$+ 0.1464466087\, \mathit{port2}^2\, \mathit{port3} + 0.2462928580\, \mathit{port1}\, \mathit{port2}^2$
$+ 0.1617785095\, \mathit{port3}^2\, \mathit{port1} + 0.3535533905\, \mathit{port2}^3 + 0.144275016\, \mathit{port1}^3$

***Sannsynligheter:***

```
> h[1]:=.60660171e-1^2;
```
\> 2 fotoner i mode 3 og 1 foton i mode 2

$$h_1 := 0.003679656346$$

```
> h[2]:=(1/2)*.492585715^2;
```
\> 1 foton i mode1, 2 og 3

$$h_2 := 0.1213203433$$

```
> h[3]:=.5000000000^2;
```
\> 2 fotoner i mode 1 og 1 i mode 2

$$h_3 := 0.2500000000$$

```
> h[4]:=.2781745915^2;
```
\> 2 fotoner i mode 1 og 1 i mode 3

$$h_4 := 0.07738110336$$

```
> h[5]:=.1464466087^2;
```
\> 2 fotoner i mode 2 og 1 i mode 3

$$h_5 := 0.02144660920$$

```
> h[6]:=(3)*.25126266e-1^2;
```
\> 3 fotoner i mode 3

$$h_6 := 0.001893987729$$

```
> h[7]:=.2462928580^2;
```
\> 1 foton i mode 1 og 2 fotoner i mode 2

$$h_7 := 0.06066017190$$

```
> h[8]:=.1617785095^2;
```
\> 2 fotoner i mode 3 og 1 i mode 1

$$h_8 := 0.02617228614$$

```
> h[9]:=(3)*.3535533905^2;
```
\> 3 fotoner i mode 2

$$h_9 := 0.3749999997$$

```
> h[10]:=(3)*.144275016^2;
```
\> 3 fotoner i mode 1

$$h_{10} := 0.06244584072$$

```
> sum('h[l]','l'=1..10);
```

$$0.9999999983$$

# Appendix B

# State change with one beamsplitter

```
> restart;
```

Try to do postcorrection with just the use of one beamsplitter

### Case 1:

**1 incoming photon:**

```
> a:=evalf(1/root[4](2)*sin(x));
```

$$a := 0.8408964155\sin(x)$$

**2 incoming photons:**

```
> b:=evalf((1-sqrt(2))*root[4](2)*sqrt(2)*sin(x)*cos(x));
```

$$b := -0.6966213991\sin(x)\cos(x)$$

**3 incoming photons:**

```
> c:=evalf((1-sqrt(2))^2/root[4](2)*sqrt(3)*sin(x)*cos(x)^2);
```

$$c := 0.2498916572\sin(x)\cos(x)^2$$

Try to solve the equations to get the non-linear sign shift with the beamsplitter:

```
> solve(a=b,x);
```

$$0., 3.141592654 + 0.6329743207\,I, 3.141592654 - 0.6329743207\,I$$

```
> solve(a=-c);
```

$$0., -1.570796327 - 1.367028558\,I, 1.570796327 + 1.367028558\,I,$$
$$-1.570796327 + 1.367028558\,I, 1.570796327 - 1.367028558\,I$$

```
> solve(b=-c);
```

$$1.570796327,\ 0.,\ -1.684518225\,I,\ 1.684518225\,I$$

Try to solve the equations just to try to get back to the original state:

```
> solve(a=b,x);
```

$$0.,\ 3.141592654 + 0.6329743207\,I,\ 3.141592654 - 0.6329743207\,I$$

```
> solve(a=c);
```

$$0.,\ 3.141592654 + 1.215587759\,I,\ -1.215587759\,I,\ 3.141592654 - 1.215587759\,I,\ 1.215587759\,I$$

```
> solve(b=c);
```

$$1.570796327,\ 0.,\ 3.141592654 + 1.684518225\,I,\ 3.141592654 - 1.684518225\,I$$

Finally a plot, to get an overlook of the different probability amplitudes:

```
> plot([a, b, c], x=0..2*Pi, color=[red,blue,green]);
```

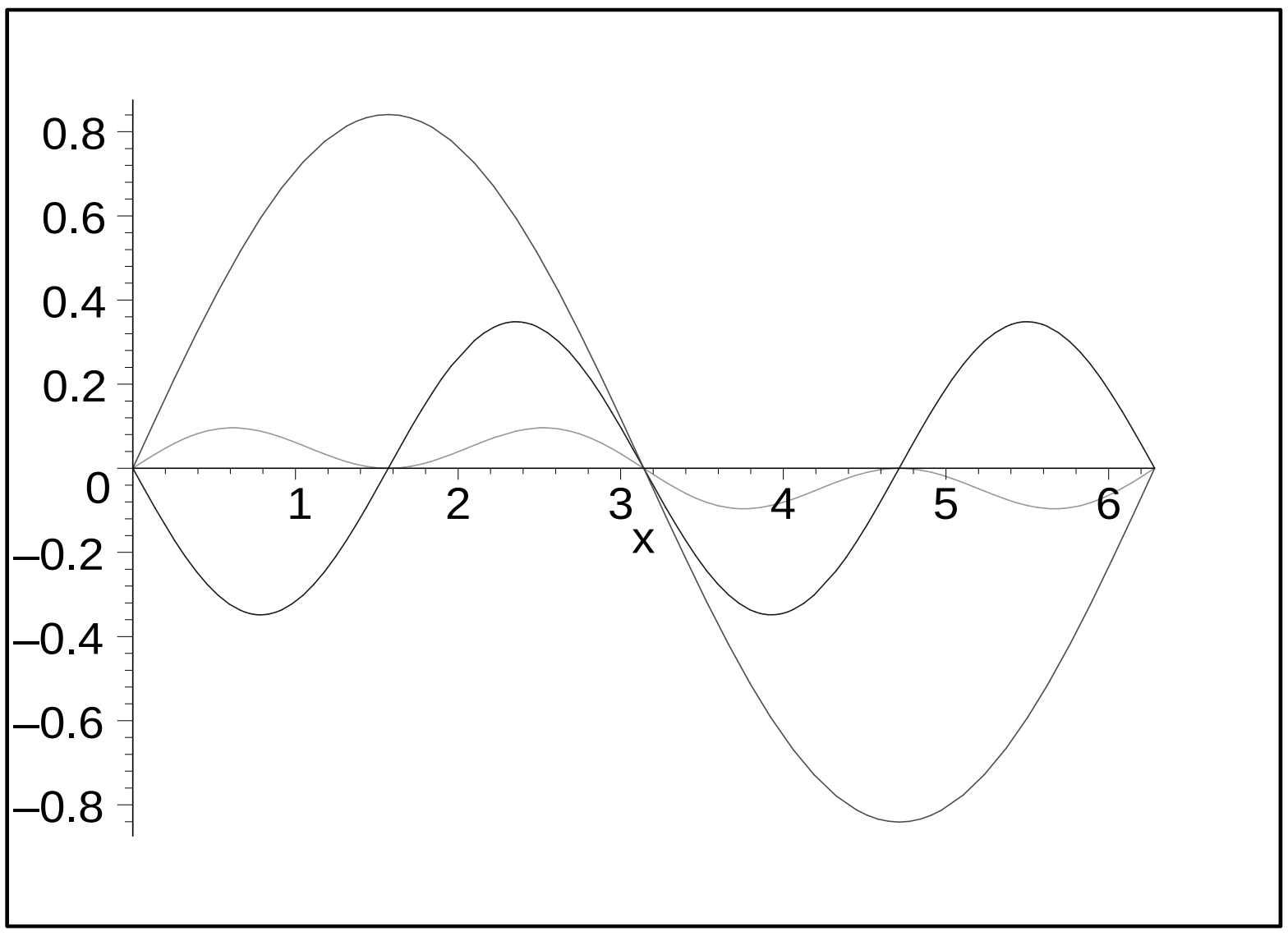

Case 3

0 incoming photons:

```
> d:=evalf(-(1/2-1/sqrt(2))*cos(y));
```

$$d := 0.2071067810\cos(y)$$

1 incoming photon:

```
> e:=evalf([(1-sqrt(2))*(1/2-1/sqrt(2))+1/root[4](2)*sqrt(3/sqrt(2)-2)]
> *(sin(y)^2-cos(y)^2));
```

$$e := [0.3786796557]\,(\sin(y)^2 - 1.\cos(y)^2)$$

**2 incoming photons:**

```
> f:=evalf([(1-sqrt(2))/root[4](2)*sqrt(3/sqrt(2)-2)+(1-sqrt(2))*[(1-sq
> rt(2))*(1/2-1/sqrt(2))+1/root[4](2)*sqrt(3/sqrt(2)-2)]]*(2*sin(y)^2*co
> s(y)-cos(y)^3));
```

$$f := [-0.1213203432 + [-0.1568542490]]\,(2.\sin(y)^2\cos(y) - 1.\cos(y)^3)$$

Try to solve the equations to get the non-linear sign shift with the beamsplitter:

```
> solve(d=e, y);
> solve(d=-f, y);
> solve(e=-f,y);
```

Try to solve the equations just to try to get back to the original state:

```
> solve(d=e, y);
> solve(d=-f, y);
> solve(e=-f, y);
```

`Don't know why, but maple won't solve this equations`

`A numerically plot of the d-,e- and f-values:`

```
> plot([.2071067810*cos(y),.3786796557*(sin(y)^2-1.*cos(y)^2),0.5563491
> 84*sin(y)^2*cos(y)-0.278174592*cos(y)^3],y=0..2*Pi,
> color=[red,blue,green]);
```

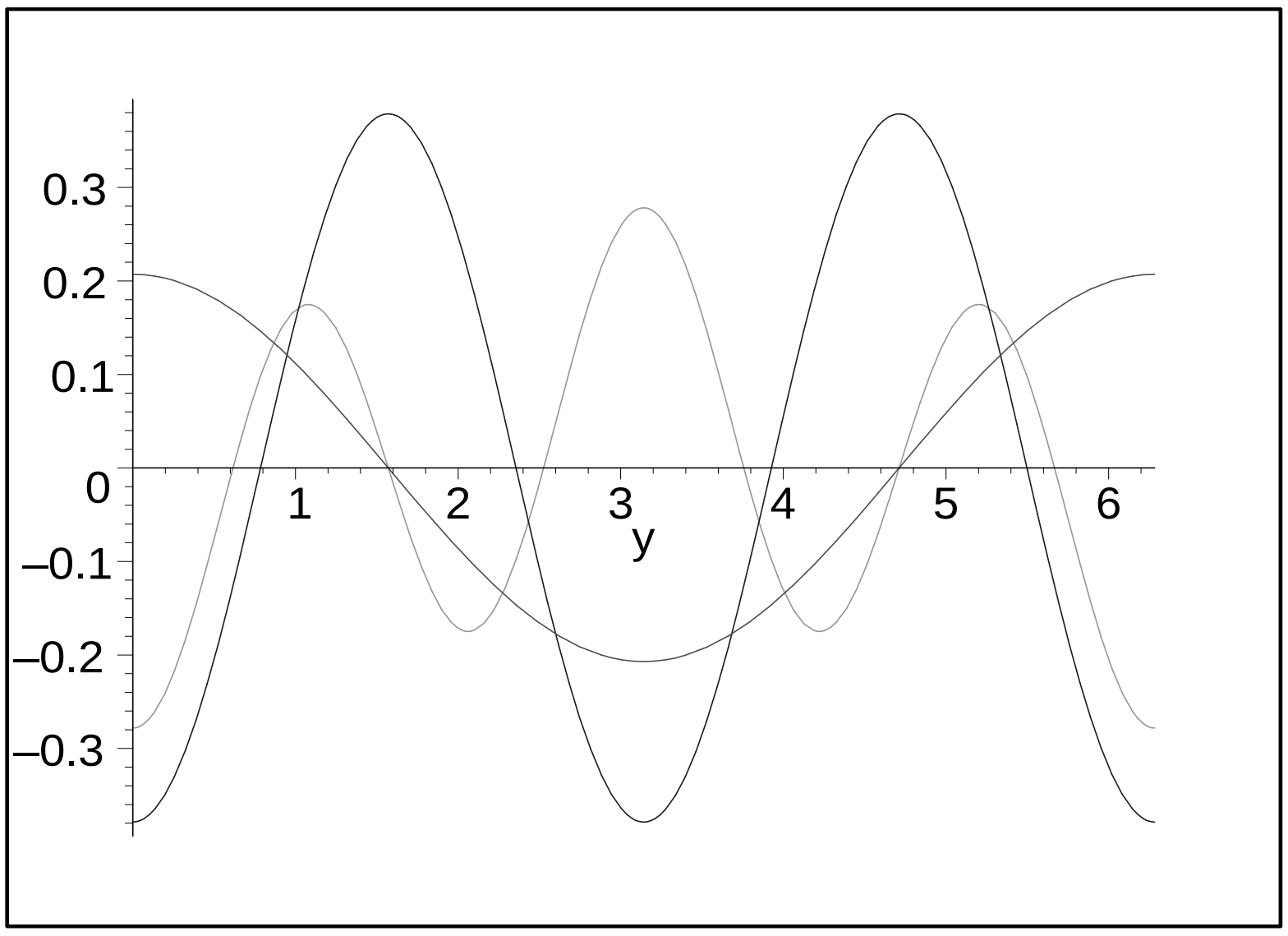

# Appendix C

# Postcorrection with two beamsplitters and phase shifter

```
> restart;
```

Case 1

The same as Case 1 in Appendix B- the sign probably doesn't matter, since we also can change the signe whit the phase shifter. All we need to do is look at the probability amplitude, and since this never equals for the 3 cases, we can't get a solution. We could have added another ancilla, but this will only complicate the calculation since we already got one additional photon in the state. In stead we choose to look at different setups.

### Case 3

0 incoming photons:

```
> d:=evalf((1/2-1/sqrt(2))*sin(x)*sin(y));
```

$$d := -0.2071067810 \sin(x)\sin(y)$$

1 incoming photon:

```
> e:=evalf([(1-sqrt(2))*(1/2-1/sqrt(2))+1/root[4](2)*sqrt(3/sqrt(2)-2)]
> *2*sin(x)*cos(x)*sin(y)*cos(y));
```

$$e := 2.\,[0.3786796557]\sin(x)\cos(x)\sin(y)\cos(y)$$

2 incoming photons:

```
> f:=evalf([(1-sqrt(2))/root[4](2)*sqrt(3/sqrt(2)-2)+(1-sqrt(2))*[(1-sq
> rt(2))*(1/2-1/sqrt(2))+1/root[4](2)*sqrt(3/sqrt(2)-2)]]*3*sin(x)*cos(x
> )^2*sin(y)*cos(y)^2);
```

$$f := 3.\,[-0.1213203432 + [-0.1568542490]]\,\sin(x)\cos(x)^2\sin(y)\cos(y)^2$$

```
> solve({d=e, d=f});
> solve({d=e, e=f});
> solve({d=f, e=f});
```

Don't know why, but maple won't solve this

To make it easier, we try to use **x=y** and se if we can get a solution with this:

```
> g:=evalf((1/2-1/sqrt(2))*sin(y)*sin(y));
```

$$g := -0.2071067810\sin(y)^2$$

```
> h:=evalf([(1-sqrt(2))*(1/2-1/sqrt(2))+1/root[4](2)*sqrt(3/sqrt(2)-2)]
> *2*sin(y)*cos(y)*sin(y)*cos(y));
```

$$h := 2.\,[0.3786796557]\sin(y)^2\cos(y)^2$$

```
> i:=evalf([(1-sqrt(2))/root[4](2)*sqrt(3/sqrt(2)-2)+(1-sqrt(2))*[(1-sq
> rt(2))*(1/2-1/sqrt(2))+1/root[4](2)*sqrt(3/sqrt(2)-2)]]*3*sin(y)*cos(y
> )^2*sin(y)*cos(y)^2);
```

$$i := 3.\,[-0.1213203432 + [-0.1568542490]]\sin(y)^2\cos(y)^4$$

```
> solve(g=h); solve(g=i); solve(h=i);
```

Don't know why, but maple won't solve this either

Finally a plot, to get an overlook of the different probability amplitudes:

```
> plot([-.2071067810*sin(y)^2,
> 0.7573593114*sin(y)^2*cos(y)^2, -.834523777*sin(y)^2*cos(y)^4],
> y=0..2*Pi, color=[red,blue,green]);
```

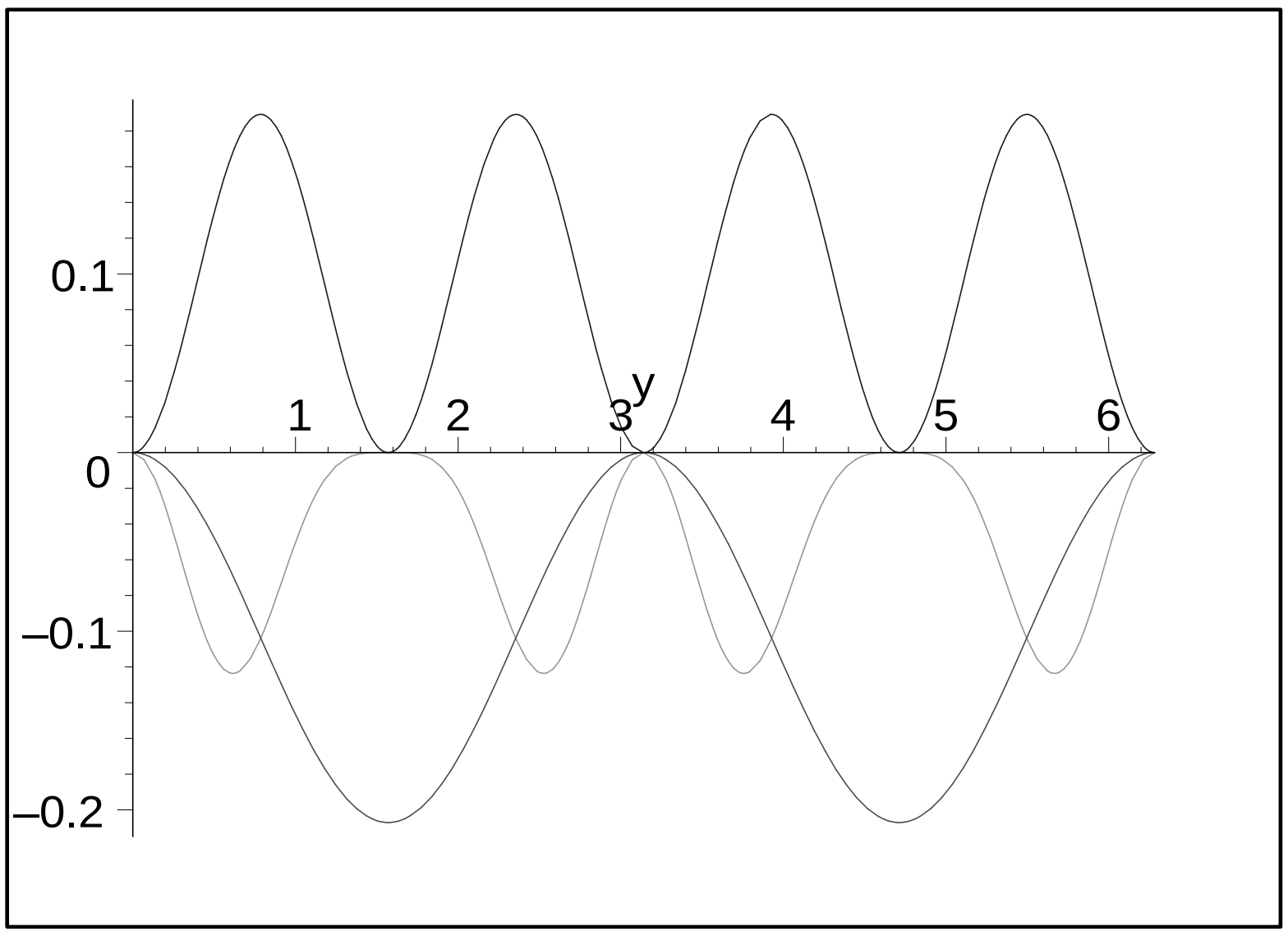

# Appendix D

# Outcomes when sending in a superposition-state

```
> restart;with(linalg):
```

Warning, the protected names norm and trace have been redefined and unprotected

An overview of possible interesting outcomes when you get inn a superposition of —1>, —2> and —3> and you wish to get a superposition of —0>, —1> and —2> out

```
> a:=port1;
> b:=port2;c:=port3;
```

$$a := \mathit{port1}$$
$$b := \mathit{port2}$$
$$c := \mathit{port3}$$

```
> u := matrix([ [-cos(t[2]),
> sin(t[2])*cos(t[3]), sin(t[2])*sin(t[3])],
> [ cos(t[1])*sin(t[2]),
> sin(t[1])*sin(t[3])+cos(t[1])*cos(t[2])*cos(t[3]),
> -sin(t[1])*cos(t[3])+cos(t[1])*cos(t[2])*sin(t[3])],
> [ sin(t[1])*sin(t[2]),
> -cos(t[1])*sin(t[3])+sin(t[1])*cos(t[2])*cos(t[3]),
> cos(t[1])*cos(t[3])+sin(t[1])*cos(t[2])*sin(t[3])]
> ]);
```

$$u := \begin{bmatrix} -\cos(t_2) & \sin(t_2)\cos(t_3) & \sin(t_2)\sin(t_3) \\ \cos(t_1)\sin(t_2) & \sin(t_1)\sin(t_3)+\cos(t_1)\cos(t_2)\cos(t_3) & -\sin(t_1)\cos(t_3)+\cos(t_1)\cos(t_2)\sin(t_3) \\ \sin(t_1)\sin(t_2) & -\cos(t_1)\sin(t_3)+\sin(t_1)\cos(t_2)\cos(t_3) & \cos(t_1)\cos(t_3)+\sin(t_1)\cos(t_2)\sin(t_3) \end{bmatrix}$$

Definerer d,e og f som rad 1,2 og 3 i u-matrisen:

```
> d:=u[1,1]*a+u[1,2]*b+u[1,3]*c;
```
> Rad 1
```
> e:=u[2,1]*a+u[2,2]*b+u[2,3]*c;
```
> Rad 2
```
> f:=u[3,1]*a+u[3,2]*b+u[3,3]*c;
```
> Rad 3

$$d := -\cos(t_2)\,\mathit{port1} + \sin(t_2)\cos(t_3)\,\mathit{port2} + \sin(t_2)\sin(t_3)\,\mathit{port3}$$

$$e := \cos(t_1)\sin(t_2)\,\mathit{port1} + (\sin(t_1)\sin(t_3) + \cos(t_1)\cos(t_2)\cos(t_3))\,\mathit{port2} + (-\sin(t_1)\cos(t_3) + \cos(t_1)\cos(t_2)\sin(t_3))\,\mathit{port3}$$

$$f := \sin(t_1)\sin(t_2)\,\mathit{port1} + (-\cos(t_1)\sin(t_3) + \sin(t_1)\cos(t_2)\cos(t_3))\,\mathit{port2} + (\cos(t_1)\cos(t_3) + \sin(t_1)\cos(t_2)\sin(t_3))\,\mathit{port3}$$

## 1 photon in the upper mode

```
> evalf(expand(d*e));
```

$$-1.\cos(t_2)\,\mathit{port1}^2\cos(t_1)\sin(t_2) - 1.\cos(t_2)\,\mathit{port1}\,\mathit{port2}\sin(t_1)\sin(t_3) - 1.\cos(t_2)^2\,\mathit{port1}\,\mathit{port2}\cos(t_1)\cos(t_3) + \cos(t_2)\,\mathit{port1}\,\mathit{port3}\sin(t_1)\cos(t_3) - 1.\cos(t_2)^2\,\mathit{port1}\,\mathit{port3}\cos(t_1)\sin(t_3) + \sin(t_2)^2\cos(t_3)\,\mathit{port2}\cos(t_1)\,\mathit{port1} + \sin(t_2)\cos(t_3)\,\mathit{port2}^2\sin(t_1)\sin(t_3) + \sin(t_2)\cos(t_3)^2\,\mathit{port2}^2\cos(t_1)\cos(t_2) - 1.\sin(t_2)\cos(t_3)^2\,\mathit{port2}\,\mathit{port3}\sin(t_1) + 2.\sin(t_2)\cos(t_3)\,\mathit{port2}\,\mathit{port3}\cos(t_1)\cos(t_2)\sin(t_3) + \sin(t_2)^2\sin(t_3)\,\mathit{port3}\cos(t_1)\,\mathit{port1} + \sin(t_2)\sin(t_3)^2\,\mathit{port3}\,\mathit{port2}\sin(t_1) - 1.\sin(t_2)\sin(t_3)\,\mathit{port3}^2\sin(t_1)\cos(t_3) + \sin(t_2)\sin(t_3)^2\,\mathit{port3}^2\cos(t_1)\cos(t_2)$$

***Possible interesting outcomes:***

```
> h[1]:=sin(t[2])*cos(t[3])*sin(t[1])*sin(t[3])+sin(t[2])*cos(t[3])^2*c
> os(t[1])*cos(t[2]);
```
> **2 photons in mode 2**

$$h_1 := \sin(t_2)\cos(t_3)\sin(t_1)\sin(t_3) + \sin(t_2)\cos(t_3)^2\cos(t_1)\cos(t_2)$$

```
> h[2]:=-1.*sin(t[2])*cos(t[3])^2*sin(t[1])+2.*sin(t[2])*cos(t[3])*cos(
> t[1])*cos(t[2])*sin(t[3]);
```
> **1 photon in mode 2 and 1 in mode 3**

$$h_2 := -1.\sin(t_2)\cos(t_3)^2\sin(t_1) + 2.\sin(t_2)\cos(t_3)\cos(t_1)\cos(t_2)\sin(t_3)$$

```
> h[3]:=-1.*sin(t[2])*sin(t[3])*port3^2*sin(t[1])*cos(t[3])+sin(t[2])*s
> in(t[3])^2*port3^2*cos(t[1])*cos(t[2]);
```
> **2 photons in mode 3**

$$h_3 := -1.\sin(t_2)\sin(t_3)\,\mathit{port3}^2\sin(t_1)\cos(t_3) + \sin(t_2)\sin(t_3)^2\,\mathit{port3}^2\cos(t_1)\cos(t_2)$$

2 photons in the upper mode

```
> evalf(expand(d^2*e));
```

$$\cos(t_2)^2\,\mathit{port1}^3\cos(t_1)\sin(t_2)+\cos(t_2)^2\,\mathit{port1}^2\,\mathit{port2}\sin(t_1)\sin(t_3)$$
$$+\cos(t_2)^3\,\mathit{port1}^2\,\mathit{port2}\cos(t_1)\cos(t_3)-1.\cos(t_2)^2\,\mathit{port1}^2\,\mathit{port3}\sin(t_1)\cos(t_3)$$
$$+\cos(t_2)^3\,\mathit{port1}^2\,\mathit{port3}\cos(t_1)\sin(t_3)-2.\cos(t_2)\,\mathit{port1}^2\sin(t_2)^2\cos(t_3)\,\mathit{port2}\cos(t_1)$$
$$-2.\cos(t_2)\,\mathit{port1}\sin(t_2)\cos(t_3)\,\mathit{port2}^2\sin(t_1)\sin(t_3)$$
$$-2.\cos(t_2)^2\,\mathit{port1}\sin(t_2)\cos(t_3)^2\,\mathit{port2}^2\cos(t_1)$$
$$+2.\cos(t_2)\,\mathit{port1}\sin(t_2)\cos(t_3)^2\,\mathit{port2}\,\mathit{port3}\sin(t_1)$$
$$-4.\cos(t_2)^2\,\mathit{port1}\sin(t_2)\cos(t_3)\,\mathit{port2}\,\mathit{port3}\cos(t_1)\sin(t_3)$$
$$-2.\cos(t_2)\,\mathit{port1}^2\sin(t_2)^2\sin(t_3)\,\mathit{port3}\cos(t_1)$$
$$-2.\cos(t_2)\,\mathit{port1}\sin(t_2)\sin(t_3)^2\,\mathit{port3}\,\mathit{port2}\sin(t_1)$$
$$+2.\cos(t_2)\,\mathit{port1}\sin(t_2)\sin(t_3)\,\mathit{port3}^2\sin(t_1)\cos(t_3)$$
$$-2.\cos(t_2)^2\,\mathit{port1}\sin(t_2)\sin(t_3)^2\,\mathit{port3}^2\cos(t_1)+\sin(t_2)^3\cos(t_3)^2\,\mathit{port2}^2\cos(t_1)\,\mathit{port1}$$
$$+\sin(t_2)^2\cos(t_3)^2\,\mathit{port2}^3\sin(t_1)\sin(t_3)+\sin(t_2)^2\cos(t_3)^3\,\mathit{port2}^3\cos(t_1)\cos(t_2)$$
$$-1.\sin(t_2)^2\cos(t_3)^3\,\mathit{port2}^2\,\mathit{port3}\sin(t_1)$$
$$+3.\sin(t_2)^2\cos(t_3)^2\,\mathit{port2}^2\,\mathit{port3}\cos(t_1)\cos(t_2)\sin(t_3)$$
$$+2.\sin(t_2)^3\cos(t_3)\,\mathit{port2}\sin(t_3)\,\mathit{port3}\cos(t_1)\,\mathit{port1}$$
$$+2.\sin(t_2)^2\cos(t_3)\,\mathit{port2}^2\sin(t_3)^2\,\mathit{port3}\sin(t_1)$$
$$-2.\sin(t_2)^2\cos(t_3)^2\,\mathit{port2}\sin(t_3)\,\mathit{port3}^2\sin(t_1)$$
$$+3.\sin(t_2)^2\cos(t_3)\,\mathit{port2}\sin(t_3)^2\,\mathit{port3}^2\cos(t_1)\cos(t_2)$$
$$+\sin(t_2)^3\sin(t_3)^2\,\mathit{port3}^2\cos(t_1)\,\mathit{port1}+\sin(t_2)^2\sin(t_3)^3\,\mathit{port3}^2\,\mathit{port2}\sin(t_1)$$
$$-1.\sin(t_2)^2\sin(t_3)^2\,\mathit{port3}^3\sin(t_1)\cos(t_3)+\sin(t_2)^2\sin(t_3)^3\,\mathit{port3}^3\cos(t_1)\cos(t_2)$$

```
{\large \textbf{\textit{Possible interesting outcomes:}}}
> i[1]:=-2.*cos(t[2])*sin(t[2])*cos(t[3])*sin(t[1])*sin(t[3])-2.*cos(t[
> 2])^2*sin(t[2])*cos(t[3])^2*cos(t[1])+sin(t[2])^3*cos(t[3])^2*cos(t[1]
> );
```

> **1 photon in mode 1 and 2 in mode 2**

$$i_1:=-2.\cos(t_2)\sin(t_2)\cos(t_3)\sin(t_1)\sin(t_3)-2.\cos(t_2)^2\sin(t_2)\cos(t_3)^2\cos(t_1)$$
$$+\sin(t_2)^3\cos(t_3)^2\cos(t_1)$$

```
> i[2]:=2.*cos(t[2])*sin(t[2])*cos(t[3])^2*sin(t[1])-4.*cos(t[2])^2*sin
> (t[2])*cos(t[3])*cos(t[1])*sin(t[3])-2.*cos(t[2])*sin(t[2])*sin(t[3])^
> 2*sin(t[1])+2.*sin(t[2])^3*cos(t[3])*sin(t[3])*cos(t[1]);
```

> **1 photon in both mode 1, 2 and 3**

$i_2 := 2.\cos(t_2)\sin(t_2)\cos(t_3)^2\sin(t_1) - 4.\cos(t_2)^2\sin(t_2)\cos(t_3)\cos(t_1)\sin(t_3)$
$- 2.\cos(t_2)\sin(t_2)\sin(t_3)^2\sin(t_1) + 2.\sin(t_2)^3\cos(t_3)\sin(t_3)\cos(t_1)$

```
> i[3]:=2.*cos(t[2])*sin(t[2])*sin(t[3])*sin(t[1])*cos(t[3])-2.*cos(t[2
> ])^2*sin(t[2])*sin(t[3])^2*cos(t[1])+sin(t[2])^3*sin(t[3])^2*cos(t[1])
> ;
```

> **1 photon in mode 1 and 2 in mode 3**

$i_3 := 2.\cos(t_2)\sin(t_2)\cos(t_3)\sin(t_1)\sin(t_3) - 2.\cos(t_2)^2\sin(t_2)\sin(t_3)^2\cos(t_1)$
$+ \sin(t_2)^3\sin(t_3)^2\cos(t_1)$

3 photons in the upper mode

```
> evalf(expand(d^3*e));
```

$$-1.\cos(t_2)^3\,\mathit{port1}^4\cos(t_1)\sin(t_2) - 1.\cos(t_2)^3\,\mathit{port1}^3\,\mathit{port2}\sin(t_1)\sin(t_3)$$
$$- 1.\cos(t_2)^4\,\mathit{port1}^3\,\mathit{port2}\cos(t_1)\cos(t_3) + \cos(t_2)^3\,\mathit{port1}^3\,\mathit{port3}\sin(t_1)\cos(t_3)$$
$$- 1.\cos(t_2)^4\,\mathit{port1}^3\,\mathit{port3}\cos(t_1)\sin(t_3)$$
$$+ 3.\cos(t_2)^2\,\mathit{port1}^3\sin(t_2)^2\cos(t_3)\,\mathit{port2}\cos(t_1)$$
$$+ 3.\cos(t_2)^2\,\mathit{port1}^2\sin(t_2)\cos(t_3)\,\mathit{port2}^2\sin(t_1)\sin(t_3)$$
$$+ 3.\cos(t_2)^3\,\mathit{port1}^2\sin(t_2)\cos(t_3)^2\,\mathit{port2}^2\cos(t_1)$$
$$- 3.\cos(t_2)^2\,\mathit{port1}^2\sin(t_2)\cos(t_3)^2\,\mathit{port2}\,\mathit{port3}\sin(t_1)$$
$$+ 6.\cos(t_2)^3\,\mathit{port1}^2\sin(t_2)\cos(t_3)\,\mathit{port2}\,\mathit{port3}\cos(t_1)\sin(t_3)$$
$$+ 3.\cos(t_2)^2\,\mathit{port1}^3\sin(t_2)^2\sin(t_3)\,\mathit{port3}\cos(t_1)$$
$$+ 3.\cos(t_2)^2\,\mathit{port1}^2\sin(t_2)\sin(t_3)^2\,\mathit{port3}\,\mathit{port2}\sin(t_1)$$
$$- 3.\cos(t_2)^2\,\mathit{port1}^2\sin(t_2)\sin(t_3)\,\mathit{port3}^2\sin(t_1)\cos(t_3)$$
$$+ 3.\cos(t_2)^3\,\mathit{port1}^2\sin(t_2)\sin(t_3)^2\,\mathit{port3}^2\cos(t_1)$$
$$- 3.\cos(t_2)\,\mathit{port1}^2\sin(t_2)^3\cos(t_3)^2\,\mathit{port2}^2\cos(t_1)$$
$$- 3.\cos(t_2)\,\mathit{port1}\sin(t_2)^2\cos(t_3)^2\,\mathit{port2}^3\sin(t_1)\sin(t_3)$$
$$- 3.\cos(t_2)^2\,\mathit{port1}\sin(t_2)^2\cos(t_3)^3\,\mathit{port2}^3\cos(t_1)$$
$$+ 3.\cos(t_2)\,\mathit{port1}\sin(t_2)^2\cos(t_3)^3\,\mathit{port2}^2\,\mathit{port3}\sin(t_1)$$
$$- 9.\cos(t_2)^2\,\mathit{port1}\sin(t_2)^2\cos(t_3)^2\,\mathit{port2}^2\,\mathit{port3}\cos(t_1)\sin(t_3)$$
$$- 6.\cos(t_2)\,\mathit{port1}^2\sin(t_2)^3\cos(t_3)\,\mathit{port2}\sin(t_3)\,\mathit{port3}\cos(t_1)$$
$$- 6.\cos(t_2)\,\mathit{port1}\sin(t_2)^2\cos(t_3)\,\mathit{port2}^2\sin(t_3)^2\,\mathit{port3}\sin(t_1)$$
$$+ 6.\cos(t_2)\,\mathit{port1}\sin(t_2)^2\cos(t_3)^2\,\mathit{port2}\sin(t_3)\,\mathit{port3}^2\sin(t_1)$$
$$- 9.\cos(t_2)^2\,\mathit{port1}\sin(t_2)^2\cos(t_3)\,\mathit{port2}\sin(t_3)^2\,\mathit{port3}^2\cos(t_1)$$
$$- 3.\cos(t_2)\,\mathit{port1}^2\sin(t_2)^3\sin(t_3)^2\,\mathit{port3}^2\cos(t_1)$$
$$- 3.\cos(t_2)\,\mathit{port1}\sin(t_2)^2\sin(t_3)^3\,\mathit{port3}^2\,\mathit{port2}\sin(t_1)$$
$$+ 3.\cos(t_2)\,\mathit{port1}\sin(t_2)^2\sin(t_3)^2\,\mathit{port3}^3\sin(t_1)\cos(t_3)$$
$$- 3.\cos(t_2)^2\,\mathit{port1}\sin(t_2)^2\sin(t_3)^3\,\mathit{port3}^3\cos(t_1) + \sin(t_2)^4\cos(t_3)^3\,\mathit{port2}^3\cos(t_1)\,\mathit{port1}$$
$$+ \sin(t_2)^3\cos(t_3)^3\,\mathit{port2}^4\sin(t_1)\sin(t_3) + \sin(t_2)^3\cos(t_3)^4\,\mathit{port2}^4\cos(t_1)\cos(t_2)$$
$$- 1.\sin(t_2)^3\cos(t_3)^4\,\mathit{port2}^3\,\mathit{port3}\sin(t_1)$$
$$+ 4.\sin(t_2)^3\cos(t_3)^3\,\mathit{port2}^3\,\mathit{port3}\cos(t_1)\cos(t_2)\sin(t_3)$$
$$+ 3.\sin(t_2)^4\cos(t_3)^2\,\mathit{port2}^2\sin(t_3)\,\mathit{port3}\cos(t_1)\,\mathit{port1}$$
$$+ 3.\sin(t_2)^3\cos(t_3)^2\,\mathit{port2}^3\sin(t_3)^2\,\mathit{port3}\sin(t_1)$$
$$- 3.\sin(t_2)^3\cos(t_3)^3\,\mathit{port2}^2\sin(t_3)\,\mathit{port3}^2\sin(t_1)$$
$$+ 6.\sin(t_2)^3\cos(t_3)^2\,\mathit{port2}^2\sin(t_3)^2\,\mathit{port3}^2\cos(t_1)\cos(t_2)$$
$$+ 3.\sin(t_2)^4\cos(t_3)\,\mathit{port2}\sin(t_3)^2\,\mathit{port3}^2\cos(t_1)\,\mathit{port1}$$
$$+ 3.\sin(t_2)^3\cos(t_3)\,\mathit{port2}^2\sin(t_3)^3\,\mathit{port3}^2\sin(t_1)$$
$$- 3.\sin(t_2)^3\cos(t_3)^2\,\mathit{port2}\sin(t_3)^2\,\mathit{port3}^3\sin(t_1)$$
$$+ 4.\sin(t_2)^3\cos(t_3)\,\mathit{port2}\sin(t_3)^3\,\mathit{port3}^3\cos(t_1)\cos(t_2)$$
$$+ \sin(t_2)^4\sin(t_3)^3\,\mathit{port3}^3\cos(t_1)\,\mathit{port1} + \sin(t_2)^3\sin(t_3)^4\,\mathit{port3}^3\,\mathit{port2}\sin(t_1)$$
$$- 1.\sin(t_2)^3\sin(t_3)^3\,\mathit{port3}^4\sin(t_1)\cos(t_3) + \sin(t_2)^3\sin(t_3)^4\,\mathit{port3}^4\cos(t_1)\cos(t_2)$$

```
> j[1]:=3.*cos(t[2])^2*sin(t[2])*cos(t[3])*sin(t[1])*sin(t[3])+3.*cos(t
> [2])^3*sin(t[2])*cos(t[3])^2*cos(t[1])-3.*cos(t[2])*sin(t[2])^3*cos(t[
> 3])^2*cos(t[1]);
```
> **2 photons in mode 1 and 2 in mode 2**

$$j_1 := 3.\cos(t_2)^2\sin(t_2)\cos(t_3)\sin(t_1)\sin(t_3) + 3.\cos(t_2)^3\sin(t_2)\cos(t_3)^2\cos(t_1) - 3.\cos(t_2)\sin(t_2)^3\cos(t_3)^2\cos(t_1)$$

```
> j[2]:=-3.*cos(t[2])^2*sin(t[2])*cos(t[3])^2*sin(t[1])+6.*cos(t[2])^3*
> sin(t[2])*cos(t[3])*cos(t[1])*sin(t[3])+3.*cos(t[2])^2*sin(t[2])*sin(t
> [3])^2*sin(t[1])+6.*cos(t[2])*sin(t[2])^3*cos(t[3])*sin(t[3])*cos(t[1]
> );
```
> **2 photons in mode 1 and 1 in mode 2 and 3**

$$j_2 := -3.\cos(t_2)^2\sin(t_2)\cos(t_3)^2\sin(t_1) + 6.\cos(t_2)^3\sin(t_2)\cos(t_3)\cos(t_1)\sin(t_3) + 3.\cos(t_2)^2\sin(t_2)\sin(t_3)^2\sin(t_1) + 6.\cos(t_2)\sin(t_2)^3\cos(t_3)\sin(t_3)\cos(t_1)$$

```
> j[3]:=-3.*cos(t[2])^2*sin(t[2])*sin(t[3])*sin(t[1])*cos(t[3])+3.*cos(
> t[2])^3*sin(t[2])*sin(t[3])^2*cos(t[1])-3.*cos(t[2])^2*sin(t[2])*sin(t
> [3])*sin(t[1])*cos(t[3])+3.*cos(t[2])^3*sin(t[2])*sin(t[3])^2*cos(t[1]
> )-3.*cos(t[2])*sin(t[2])^3*sin(t[3])^2*cos(t[1]);
```
> **2 photons in mode 1 and 2 in mode 3**

$$j_3 := -6.\cos(t_2)^2\sin(t_2)\cos(t_3)\sin(t_1)\sin(t_3) + 6.\cos(t_2)^3\sin(t_2)\sin(t_3)^2\cos(t_1) - 3.\cos(t_2)\sin(t_2)^3\sin(t_3)^2\cos(t_1)$$

# Appendix E

# Postcorrection in a modified non-linear sign shift-gate

```
> restart;
```

Case 1

Try to send this state to another NS-gate to execute the postcorrection there.

1 incoming photon:

```
> a:=1/root[4](2);
```

$$a := \frac{2^{(3/4)}}{2}$$

2 incoming photons:

```
> b:=(1-sqrt(2))*root[4](2);
```

$$b := (1 - \sqrt{2})\, 2^{(1/4)}$$

3 incoming photons:

```
> c:=(1-sqrt(2))^2*1/root[4](2);
```

$$c := \frac{(1 - \sqrt{2})^2\, 2^{(3/4)}}{2}$$

If we assume we get **a—1>+b—2>+c—3>** in, we wish to get **a*a1—0>+b*b1—1>-c*c1—2>** out, where **—a*a1—=—b*b1—=—c*c1—** to get the postcorrection we seek (the ratio between the states must be the same as in the original state). For simplicity lets say we detect two photons in mode 2.

The probabilityamplitude for the correction of 1 incoming photon:

```
> a1:=sin(t[2])*cos(t[3])*sin(t[1])*sin(t[3])+sin(t[2])*cos(t[3])^2*cos
> (t[1])*cos(t[2]);
```

$$a1 := \sin(t_2)\cos(t_3)\sin(t_1)\sin(t_3) + \sin(t_2)\cos(t_3)^2\cos(t_1)\cos(t_2)$$

The probabilityamplitude for the correction of 2 incoming photons:

```
> b1:=-2.*cos(t[2])*sin(t[2])*cos(t[3])*sin(t[1])*sin(t[3])-2.*cos(t[2]
> )^2*sin(t[2])*cos(t[3])^2*cos(t[1])+sin(t[2])^3*cos(t[3])^2*cos(t[1]);
```

$$b1 := -2.\cos(t_2)\sin(t_2)\cos(t_3)\sin(t_1)\sin(t_3) - 2.\cos(t_2)^2\sin(t_2)\cos(t_3)^2\cos(t_1) + \sin(t_2)^3\cos(t_3)^2\cos(t_1)$$

The probabilityamplitude for the correction of 3 incoming photons:

```
> c1:=3.*cos(t[2])^2*sin(t[2])*cos(t[3])*sin(t[1])*sin(t[3])+3.*cos(t[2
> ])^3*sin(t[2])*cos(t[3])^2*cos(t[1])-3.*cos(t[2])*sin(t[2])^3*cos(t[3]
> )^2*cos(t[1]);
```

$$c1 := 3.\cos(t_2)^2\sin(t_2)\cos(t_3)\sin(t_1)\sin(t_3) + 3.\cos(t_2)^3\sin(t_2)\cos(t_3)^2\cos(t_1) - 3.\cos(t_2)\sin(t_2)^3\cos(t_3)^2\cos(t_1)$$

```
> solve( {a*a1=b*b1,a*a1=-c*c1,b*b1=-c*c1 } );
```

$$\{t_3 = 1.570796327,\ t_1 = t_1,\ t_2 = t_2\},\ \{t_2 = 0.,\ t_1 = t_1,\ t_3 = t_3\},$$
$$\{t_2 = -2.466864691,\ t_3 = -1.\arctan(-\frac{0.6614985514}{\tan(t_1)}),\ t_1 = t_1\},$$
$$\{t_2 = 2.466864691,\ t_3 = -1.\arctan(-\frac{0.6614985514}{\tan(t_1)}),\ t_1 = t_1\},$$
$$\{t_2 = -1.561539827\,I,\ t_3 = -1.\arctan(\frac{4.075712114}{\tan(t_1)}),\ t_1 = t_1\},$$
$$\{t_2 = 1.561539827\,I,\ t_3 = -1.\arctan(\frac{4.075712114}{\tan(t_1)}),\ t_1 = t_1\}$$

Try to evaluate each of the probability amplitudes, to see if they equal each other.

```
> eval(a*a1, [t[2]=-2.466864691,
> t[3]=-1.*arctan(-.6614985514/tan(2)),
> t[1]=2]);
```

$$\frac{1}{2}2^{(3/4)}(-0.6246849121\cos(1.\arctan(\frac{0.6614985514}{\tan(2)}))\sin(2)\sin(1.\arctan(\frac{0.6614985514}{\tan(2)})) + 0.4878020287\cos(1.\arctan(\frac{0.6614985514}{\tan(2)}))^2\cos(2))$$

```
> eval(b*b1, [t[2]=-2.466864691,
> t[3]=-1.*arctan(-.6614985514/tan(2)),
> t[1]=2]);
```

$$(1-\sqrt{2})\,2^{(1/4)}($$
$$-0.9756040574\cos(1.\arctan(\frac{0.6614985514}{\tan(2)}))\sin(2)\sin(1.\arctan(\frac{0.6614985514}{\tan(2)}))$$
$$+\,0.5180551217\cos(1.\arctan(\frac{0.6614985514}{\tan(2)}))^2\cos(2))$$

```
> eval(c*c1, [t[2]=-2.466864691,
> t[3]=-1.*arctan(-.6614985514/tan(2)),
> t[1]=2]);
```

$$\frac{1}{2}(1-\sqrt{2})^2\,2^{(3/4)}($$
$$-1.142740034\cos(1.\arctan(\frac{0.6614985514}{\tan(2)}))\sin(2)\sin(1.\arctan(\frac{0.6614985514}{\tan(2)}))$$
$$+\,0.3212725446\cos(1.\arctan(\frac{0.6614985514}{\tan(2)}))^2\cos(2))$$

## Case 3

0 incoming photons:

```
> d:=1/2-1/sqrt(2);
```

$$d := \frac{1}{2}-\frac{\sqrt{2}}{2}$$

1 incoming photon:

```
> e:=(1-sqrt(2))*(1/2-1/sqrt(2))+1/root[4](2)*sqrt(3/sqrt(2)-2);
```

$$e := (1-\sqrt{2})\,(\frac{1}{2}-\frac{\sqrt{2}}{2})+\frac{2^{(3/4)}\,\sqrt{-8+6\sqrt{2}}}{4}$$

2 incoming photons:

```
> f:=(1-sqrt(2))/root[4](2)*sqrt(3/sqrt(2)-2)+(1-sqrt(2))*[(1-sqrt(2))*
> (1/2-1/sqrt(2))+1/root[4](2)*sqrt(3/sqrt(2)-2)];
```

$$f := \frac{(1-\sqrt{2})\,2^{(3/4)}\,\sqrt{-8+6\sqrt{2}}}{4}+(1-\sqrt{2})\,[(1-\sqrt{2})\,(\frac{1}{2}-\frac{\sqrt{2}}{2})+\frac{2^{(3/4)}\,\sqrt{-8+6\sqrt{2}}}{4}]$$

If we assume we get **-—0>+—1>-—2>** in, we wish to get **d1—0>+e1—1>-f1—2>** out, where **—d*d1—=—e*e1—=—f*f1—** to get the postcorrection we seek (the ratio between the states must be the same as in the original state). For simplicity lets say we detect one photon in mode 2 in the NS-gate.

The probabilityamplitude for the correction of 0 incoming photons:

```
> d1:=sin(t[1])*sin(t[3])+cos(t[1])*cos(t[2])*cos(t[3]);
```

$$d1 := \sin(t_1)\sin(t_3) + \cos(t_1)\cos(t_2)\cos(t_3)$$

The probabilityamplitude for the correction of 1 incoming photon:

```
> e1:=-sin(t[1])*cos(t[2])*sin(t[3])-cos(t[1])*cos(t[2])^2*cos(t[3])+co
> s(t[1])*sin(t[2])^2*cos(t[3]);
```

$$e1 := -\cos(t_2)\sin(t_1)\sin(t_3) - \cos(t_2)^2\cos(t_3)\cos(t_1) + \sin(t_2)^2\cos(t_3)\cos(t_1)$$

The probabilityamplitude for the correction of 2 incoming photons:

```
> f1:=sin(t[1])*cos(t[2])^2*sin(t[3])+cos(t[1])*cos(t[2])^3*cos(t[3])-2
> .*cos(t[1])*cos(t[2])*sin(t[2])^2*cos(t[3]);
```

$$f1 := \cos(t_2)^2\sin(t_1)\sin(t_3) + \cos(t_2)^3\cos(t_3)\cos(t_1) - 2.\cos(t_2)\sin(t_2)^2\cos(t_3)\cos(t_1)$$

```
> solve( {d*d1=e*e1, d*d1=-f*f1, e*e1=-f*f1} );
```

# Appendix F

# General Non-linear sign shift gate calculations

```
> restart;with(linalg):
```

```
Warning, the protected names norm and trace have been redefined
and unprotected
```

```
> a:=port1;
> b:=port2;c:=port3;
```

$$a := port1$$
$$b := port2$$
$$c := port3$$

```
> u := matrix([ [-cos(t[2]),
> sin(t[2])*cos(t[3]), sin(t[2])*sin(t[3])],
> [ cos(t[1])*sin(t[2]),
> sin(t[1])*sin(t[3])+cos(t[1])*cos(t[2])*cos(t[3]),
> -sin(t[1])*cos(t[3])+cos(t[1])*cos(t[2])*sin(t[3])],
> [ sin(t[1])*sin(t[2]),
> -cos(t[1])*sin(t[3])+sin(t[1])*cos(t[2])*cos(t[3]),
> cos(t[1])*cos(t[3])+sin(t[1])*cos(t[2])*sin(t[3])]
> ]);
```

$u :=$

$[-\cos(t_2)\,,\ \sin(t_2)\cos(t_3)\,,\ \sin(t_2)\sin(t_3)]$

$\Big[\cos(t_1)\sin(t_2)\,,\ \sin(t_1)\sin(t_3) + \cos(t_1)\cos(t_2)\cos(t_3)\,,$

$-\sin(t_1)\cos(t_3) + \cos(t_1)\cos(t_2)\sin(t_3)\Big]$

$\Big[\sin(t_1)\sin(t_2)\,,\ -\cos(t_1)\sin(t_3) + \sin(t_1)\cos(t_2)\cos(t_3)\,,$

$\cos(t_1)\cos(t_3) + \sin(t_1)\cos(t_2)\sin(t_3)\Big]$

Definerer d,e og f som rad 1,2 og 3 i u-matrisen:

```
> d:=u[1,1]*a+u[1,2]*b+u[1,3]*c;
```
> Rad 1
```
> e:=u[2,1]*a+u[2,2]*b+u[2,3]*c;
```
> Rad 2
```
> f:=u[3,1]*a+u[3,2]*b+u[3,3]*c;
```
> Rad 3

$$d := -\cos(t_2)\, \mathit{port1} + \sin(t_2)\cos(t_3)\, \mathit{port2} + \sin(t_2)\sin(t_3)\, \mathit{port3}$$

$$e := \cos(t_1)\sin(t_2)\, \mathit{port1} + (\sin(t_1)\sin(t_3) + \cos(t_1)\cos(t_2)\cos(t_3))\, \mathit{port2} + (-\sin(t_1)\cos(t_3) + \cos(t_1)\cos(t_2)\sin(t_3))\, \mathit{port3}$$

$$f := \sin(t_1)\sin(t_2)\, \mathit{port1} + (-\cos(t_1)\sin(t_3) + \sin(t_1)\cos(t_2)\cos(t_3))\, \mathit{port2} + (\cos(t_1)\cos(t_3) + \sin(t_1)\cos(t_2)\sin(t_3))\, \mathit{port3}$$

## 0 photons in the upper mode:

```
> evalf(expand(e));
```

$$\cos(t_1)\sin(t_2)\, \mathit{port1} + \mathit{port2}\sin(t_1)\sin(t_3) + \mathit{port2}\cos(t_1)\cos(t_2)\cos(t_3) - 1.\, \mathit{port3}\sin(t_1)\cos(t_3) + \mathit{port3}\cos(t_1)\cos(t_2)\sin(t_3)$$

### *Interesting outcomes:*

```
> g[1]:=-cos(t[1])*sin(t[2]);
```
> **1 photon in port 1**

$$g_1 := -\cos(t_1)\sin(t_2)$$

```
> g[2]:=sin(t[1])*sin(t[3])+cos(t[1])*cos(t[2])*cos(t[3]);
```
> **1 photon in port 2**

$$g_2 := \sin(t_1)\sin(t_3) + \cos(t_1)\cos(t_2)\cos(t_3)$$

```
> g[3]:=sin(t[1])*cos(t[3])-cos(t[1])*cos(t[2])*sin(t[3]);
```
> **1 photon in port 3**

$$g_3 := \sin(t_1)\cos(t_3) - \cos(t_1)\cos(t_2)\sin(t_3)$$

## 1 photon in the upper mode

```
> evalf(expand(d*e));
```

$$-1.\cos(t_2)\,\mathit{port1}^2\cos(t_1)\sin(t_2) - 1.\cos(t_2)\,\mathit{port1}\,\mathit{port2}\sin(t_1)\sin(t_3)$$
$$- 1.\cos(t_2)^2\,\mathit{port1}\,\mathit{port2}\cos(t_1)\cos(t_3) + \cos(t_2)\,\mathit{port1}\,\mathit{port3}\sin(t_1)\cos(t_3)$$
$$- 1.\cos(t_2)^2\,\mathit{port1}\,\mathit{port3}\cos(t_1)\sin(t_3) + \sin(t_2)^2\cos(t_3)\,\mathit{port2}\cos(t_1)\,\mathit{port1}$$
$$+ \sin(t_2)\cos(t_3)\,\mathit{port2}^2\sin(t_1)\sin(t_3) + \sin(t_2)\cos(t_3)^2\,\mathit{port2}^2\cos(t_1)\cos(t_2)$$
$$- 1.\sin(t_2)\cos(t_3)^2\,\mathit{port2}\,\mathit{port3}\sin(t_1)$$
$$+ 2.\sin(t_2)\cos(t_3)\,\mathit{port2}\,\mathit{port3}\cos(t_1)\cos(t_2)\sin(t_3)$$
$$+ \sin(t_2)^2\sin(t_3)\,\mathit{port3}\cos(t_1)\,\mathit{port1} + \sin(t_2)\sin(t_3)^2\,\mathit{port3}\,\mathit{port2}\sin(t_1)$$
$$- 1.\sin(t_2)\sin(t_3)\,\mathit{port3}^2\sin(t_1)\cos(t_3) + \sin(t_2)\sin(t_3)^2\,\mathit{port3}^2\cos(t_1)\cos(t_2)$$

***Interesting outcomes:***

```
> h[1]:=sqrt(2)*cos(t[1])*cos(t[2])*sin(t[2]);
```

> **2 photones in mode 1**

$$h_1 := \sqrt{2}\cos(t_1)\cos(t_2)\sin(t_2)$$

```
> h[2]:=-sin(t[1])*cos(t[2])*sin(t[3])-cos(t[1])*cos(t[2])^2*cos(t[3])+
> cos(t[1])*sin(t[2])^2*cos(t[3]);
```

> **1 photon in mode 1 and 1 in mode 2**

$$h_2 := -\sin(t_1)\cos(t_2)\sin(t_3) - \cos(t_1)\cos(t_2)^2\cos(t_3) + \cos(t_1)\sin(t_2)^2\cos(t_3)$$

```
> h[3]:=-sin(t[1])*cos(t[2])*cos(t[3])+cos(t[1])*cos(t[2])^2*sin(t[3])-
> cos(t[1])*sin(t[2])^2*sin(t[3]);
```

> **1 photon in mode 1 and 1 in mode 3**

$$h_3 := -\sin(t_1)\cos(t_2)\cos(t_3) + \cos(t_1)\cos(t_2)^2\sin(t_3) - \cos(t_1)\sin(t_2)^2\sin(t_3)$$

2 photons in the upper mode

```
> evalf(expand(d^2*e));
```

$$
\begin{aligned}
&\cos(t_2)^2\, \mathit{port1}^3 \cos(t_1)\sin(t_2) + \cos(t_2)^2\, \mathit{port1}^2\, \mathit{port2} \sin(t_1)\sin(t_3)\\
&+ \cos(t_2)^3\, \mathit{port1}^2\, \mathit{port2} \cos(t_1)\cos(t_3) - 1.\cos(t_2)^2\, \mathit{port1}^2\, \mathit{port3} \sin(t_1)\cos(t_3)\\
&+ \cos(t_2)^3\, \mathit{port1}^2\, \mathit{port3} \cos(t_1)\sin(t_3)\\
&- 2.\cos(t_2)\, \mathit{port1}^2 \sin(t_2)^2\cos(t_3)\, \mathit{port2} \cos(t_1)\\
&- 2.\cos(t_2)\, \mathit{port1} \sin(t_2)\cos(t_3)\, \mathit{port2}^2 \sin(t_1)\sin(t_3)\\
&- 2.\cos(t_2)^2\, \mathit{port1} \sin(t_2)\cos(t_3)^2\, \mathit{port2}^2 \cos(t_1)\\
&+ 2.\cos(t_2)\, \mathit{port1} \sin(t_2)\cos(t_3)^2\, \mathit{port2}\, \mathit{port3} \sin(t_1)\\
&- 4.\cos(t_2)^2\, \mathit{port1} \sin(t_2)\cos(t_3)\, \mathit{port2}\, \mathit{port3} \cos(t_1)\sin(t_3)\\
&- 2.\cos(t_2)\, \mathit{port1}^2 \sin(t_2)^2\sin(t_3)\, \mathit{port3} \cos(t_1)\\
&- 2.\cos(t_2)\, \mathit{port1} \sin(t_2)\sin(t_3)^2\, \mathit{port3}\, \mathit{port2} \sin(t_1)\\
&+ 2.\cos(t_2)\, \mathit{port1} \sin(t_2)\sin(t_3)\, \mathit{port3}^2 \sin(t_1)\cos(t_3)\\
&- 2.\cos(t_2)^2\, \mathit{port1} \sin(t_2)\sin(t_3)^2\, \mathit{port3}^2 \cos(t_1)\\
&+ \sin(t_2)^3\cos(t_3)^2\, \mathit{port2}^2 \cos(t_1)\, \mathit{port1} + \sin(t_2)^2\cos(t_3)^2\, \mathit{port2}^3 \sin(t_1)\sin(t_3)\\
&+ \sin(t_2)^2\cos(t_3)^3\, \mathit{port2}^3 \cos(t_1)\cos(t_2) - 1.\sin(t_2)^2\cos(t_3)^3\, \mathit{port2}^2\, \mathit{port3} \sin(t_1)\\
&+ 3.\sin(t_2)^2\cos(t_3)^2\, \mathit{port2}^2\, \mathit{port3} \cos(t_1)\cos(t_2)\sin(t_3)\\
&+ 2.\sin(t_2)^3\cos(t_3)\, \mathit{port2} \sin(t_3)\, \mathit{port3} \cos(t_1)\, \mathit{port1}\\
&+ 2.\sin(t_2)^2\cos(t_3)\, \mathit{port2}^2 \sin(t_3)^2\, \mathit{port3} \sin(t_1)\\
&- 2.\sin(t_2)^2\cos(t_3)^2\, \mathit{port2} \sin(t_3)\, \mathit{port3}^2 \sin(t_1)\\
&+ 3.\sin(t_2)^2\cos(t_3)\, \mathit{port2} \sin(t_3)^2\, \mathit{port3}^2 \cos(t_1)\cos(t_2)\\
&+ \sin(t_2)^3\sin(t_3)^2\, \mathit{port3}^2 \cos(t_1)\, \mathit{port1} + \sin(t_2)^2\sin(t_3)^3\, \mathit{port3}^2\, \mathit{port2} \sin(t_1)\\
&- 1.\sin(t_2)^2\sin(t_3)^2\, \mathit{port3}^3 \sin(t_1)\cos(t_3) + \sin(t_2)^2\sin(t_3)^3\, \mathit{port3}^3 \cos(t_1)\cos(t_2)
\end{aligned}
$$

```
{\large \textbf{\textit{Interesting outcomes:}}}
> i[1]:=-sqrt(3)*cos(t[1])*cos(t[2])^2*sin(t[2]);
```

> **3 photons in mode 1**

$$i_1 := -\sqrt{3}\cos(t_1)\cos(t_2)^2\sin(t_2)$$

```
> i[2]:=sin(t[1])*cos(t[2])^2*sin(t[3])+cos(t[1])*cos(t[2])^3*cos(t[3])
> -2.*cos(t[1])*cos(t[2])*sin(t[2])^2*cos(t[3]);
```

> **2 photons in mode 1 and 1 in mode 2**

$$i_2 := \sin(t_1)\cos(t_2)^2\sin(t_3) + \cos(t_1)\cos(t_2)^3\cos(t_3) - 2.\cos(t_1)\cos(t_2)\sin(t_2)^2\cos(t_3)$$

```
> i[3]:=-cos(t[1])*cos(t[2])^3*sin(t[3])+sin(t[1])*cos(t[2])^2*cos(t[3]
> )+2.*cos(t[1])*cos(t[2])*sin(t[2])^2*sin(t[3]);
```

> **2 photons in mode 1 and 1 in mode 3**

$$i_3 := -\cos(t_1)\cos(t_2)^3\sin(t_3) + \sin(t_1)\cos(t_2)^2\cos(t_3) + 2.\cos(t_1)\cos(t_2)\sin(t_2)^2\sin(t_3)$$

**The derived of each:**

***0 photons in:***

```
> j[1]:=diff(g[1],t[1],t[2]);
```

$$j_1 := \sin(t_1)\cos(t_2)$$

```
> j[2]:=diff(g[2],t[1],t[2],t[3]);
```

$$j_2 := -\sin(t_1)\sin(t_2)\sin(t_3)$$

```
> j[3]:=diff(g[3],t[1],t[2],t[3]);
```

$$j_3 := -\sin(t_1)\sin(t_2)\cos(t_3)$$

***1 photon in:***

```
> k[1]:=diff(h[1],t[1],t[2]);
```

$$k_1 := \sqrt{2}\sin(t_1)\sin(t_2)^2 - \sqrt{2}\sin(t_1)\cos(t_2)^2$$

```
> k[2]:=diff(h[2],t[1],t[2],t[3]);
```

$$k_2 := \cos(t_1)\sin(t_2)\cos(t_3) + 4\sin(t_1)\cos(t_2)\sin(t_3)\sin(t_2)$$

```
> k[3]:=diff(h[3],t[1],t[2],t[3]);
```

$$k_3 := -\cos(t_1)\sin(t_2)\sin(t_3) + 4\sin(t_1)\cos(t_2)\cos(t_3)\sin(t_2)$$

***2 photons in:***

```
> l[1]:=diff(i[1],t[1],t[2]);
```

$$l_1 := -2\sqrt{3}\sin(t_1)\cos(t_2)\sin(t_2)^2 + \sqrt{3}\sin(t_1)\cos(t_2)^3$$

```
> l[2]:=diff(i[2],t[1],t[2],t[3]);
```

$$l_2 := -2\cos(t_1)\cos(t_2)\cos(t_3)\sin(t_2) - 7.\sin(t_1)\cos(t_2)^2\sin(t_3)\sin(t_2) + 2.\sin(t_1)\sin(t_2)^3\sin(t_3)$$

```
> l[3]:=diff(i[3],t[1],t[2],t[3]);
```

$$l_3 := 2\cos(t_1)\cos(t_2)\sin(t_3)\sin(t_2) - 7.\sin(t_1)\cos(t_2)^2\cos(t_3)\sin(t_2) + 2.\sin(t_1)\sin(t_2)^3\cos(t_3)$$

```
> eqns:= {g[2]>0, h[2]>0, i[2]<0};
```

$$eqns := \{0 < \sin(t_1)\sin(t_3) + \cos(t_1)\cos(t_2)\cos(t_3),$$
$$0 < -\sin(t_1)\cos(t_2)\sin(t_3) - \cos(t_1)\cos(t_2)^2\cos(t_3) + \cos(t_1)\sin(t_2)^2\cos(t_3),$$
$$\sin(t_1)\cos(t_2)^2\sin(t_3) + \cos(t_1)\cos(t_2)^3\cos(t_3) - 2.\cos(t_1)\cos(t_2)\sin(t_2)^2\cos(t_3) < 0$$
$$\}$$

```
> solve( eqns, {t[1],t[2],t[3]} );
```

```
> eqn1:= {j[1]=0, k[1]=0, l[1]=0};
```

$$eqn1 := \{\sin(t_1)\cos(t_2) = 0,\ \sqrt{2}\sin(t_1)\sin(t_2)^2 - \sqrt{2}\sin(t_1)\cos(t_2)^2 = 0,$$
$$-2\sqrt{3}\sin(t_1)\cos(t_2)\sin(t_2)^2 + \sqrt{3}\sin(t_1)\cos(t_2)^3 = 0\}$$

```
> soleq:= solve(eqn1);
```

$$soleq := \{t_1 = 0,\ t_2 = t_2\}$$

```
> eqn2:= {j[2]=0, k[2]=0, l[2]=0};
```

$$eqn2 := \{-\sin(t_1)\sin(t_2)\sin(t_3) = 0,$$
$$\cos(t_1)\sin(t_2)\cos(t_3) + 4\sin(t_1)\cos(t_2)\sin(t_3)\sin(t_2) = 0,$$
$$-2\cos(t_1)\cos(t_2)\cos(t_3)\sin(t_2) - 7.\sin(t_1)\cos(t_2)^2\sin(t_3)\sin(t_2)$$
$$+ 2.\sin(t_1)\sin(t_2)^3\sin(t_3) = 0\}$$

```
> eqn3:= {j[3]=0, k[3]=0, l[3]=0};
```

$$eqn3 := \{-\sin(t_1)\sin(t_2)\cos(t_3) = 0,$$
$$-\cos(t_1)\sin(t_2)\sin(t_3) + 4\sin(t_1)\cos(t_2)\cos(t_3)\sin(t_2) = 0,$$
$$2\cos(t_1)\cos(t_2)\sin(t_3)\sin(t_2) - 7.\sin(t_1)\cos(t_2)^2\cos(t_3)\sin(t_2)$$
$$+ 2.\sin(t_1)\sin(t_2)^3\cos(t_3) = 0\}$$

A plot to find the greatest probabilitie amplitudes, with the right sign.

```
> t[3]=Pi*65/180:
> plot3d({g[1]},t[1]=0..2*Pi,t[2]=0..10,grid=[25,15],style=patch);
```

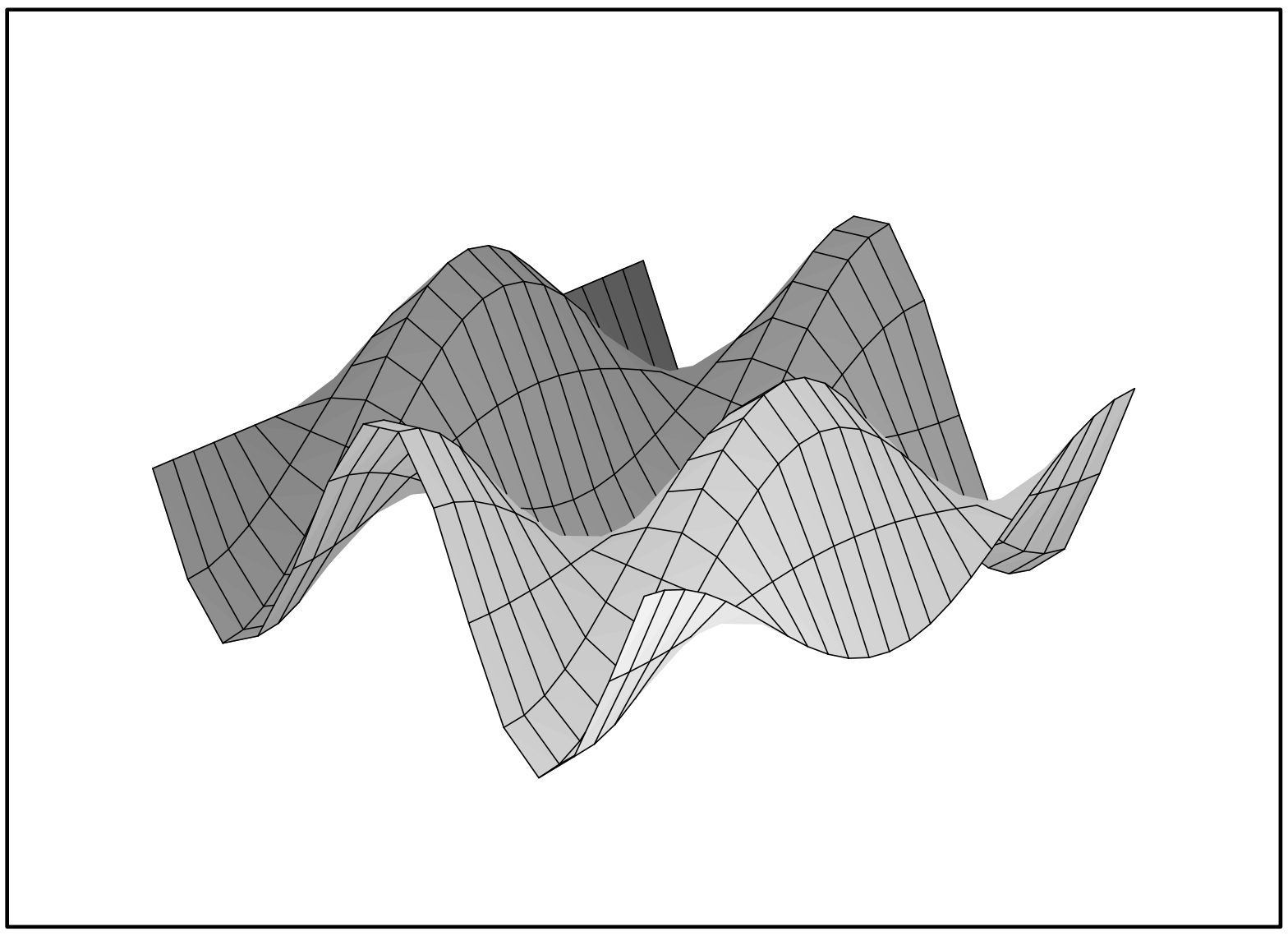

The probability amplitudes and the probability in case 1:

```
> t[1]:=3.2:t[2]:=Pi*60/180:  t[3]:=t[1]:
> evalf({g[1],g[1]^2});evalf({h[1],h[1]^2});evalf({i[1],i[1]^2});
```

$$\{0.7474443444, 0.8645486365\}$$
$$\{0.3737221725, -0.6113282036\}$$
$$\{0.1401458147, 0.3743605410\}$$

```
> evalf({g[2],g[2]^2});evalf({h[2],h[2]^2});evalf({i[2],i[2]^2});
```

$$\{0.5017037703, 0.2517066731\}$$
$$\{0.4965924595, 0.2466040708\}$$
$$\{-0.6220184020, 0.3869068924\}$$

```
> evalf({g[3],g[3]^2});evalf({h[3],h[3]^2});evalf({i[3],i[3]^2});
```

$$\{0.02913730122, 0.0008489823224\}$$
$$\{-0.05827460244, 0.003395929290\}$$
$$\{0.05099027712, 0.002600008361\}$$

The probability amplitudes and the probability in case 2:

```
> t[1]:=0.09:t[2]:=Pi*80/180:  t[3]:=t[1]:
> evalf({g[1],g[1]^2});evalf({h[1],h[1]^2});evalf({i[1],i[1]^2});
```

$$\{-0.9808219732, 0.9620117431\}$$

$$\{0.2408659518, 0.05801640676\}$$

$$\{-0.05122609757, 0.002624113071\}$$

```
> evalf({g[2],g[2]^2});evalf({h[2],h[2]^2});evalf({i[2],i[2]^2});
```

$$\{0.1803235743, 0.03251659145\}$$

$$\{0.9306988831, 0.8662004110\}$$

$$\{-0.3286657504, 0.1080211755\}$$

```
> evalf({g[3],g[3]^2});evalf({h[3],h[3]^2});evalf({i[3],i[3]^2});
```

$$\{0.07397070716, 0.005471665518\}$$

$$\{-0.09966046410, 0.009932208105\}$$

$$\{0.03238122614, 0.001048543806\}$$